\documentstyle[twoside]{article}

\makeatletter

\@addtoreset{equation}{section}   

\@addtoreset{figure}{section}
\def\thefigure{\thesection.\@arabic\c@figure}
\@addtoreset{table}{section}
\def\thetable{\thesection.\@arabic\c@table}

\def\sameauthor{\leavevmode\vrule height 2pt depth -1.6pt width 23pt}

\def\@biblabel#1{}
\def\thebibliography#1{\par\addvspace{.25in}%
\begin{center}\footnotesize REFERENCES\end{center}\@mkboth
 {REFERENCES}{REFERENCES}\addvspace{.15in}\nopagebreak\list
 {[\arabic{enumi}]}{\settowidth\labelwidth{}\leftmargin\labelwidth
 \labelsep 0pt
 \advance\leftmargin\labelsep \advance\leftmargin 16pt \itemindent -16pt
 \usecounter{enumi}}
 \def\newblock{\hskip .11em plus .33em minus .07em}
 \sloppy\clubpenalty4000\widowpenalty4000
 \sfcode`\.=1000\relax\footnotesize}
\makeatother

\newcommand{\tab}{\>}

\renewcommand{\>}{\right\rangle}
\renewcommand{\|}{\left|}

\newcommand{\mod}[1]{{\rm \ (mod\ }#1)}
\renewcommand{\bmod}[1]{{\rm (mod\ }#1)}

\newcommand{\lr}[2]{{\{#1\}_{#2}}}

\begin{document}

\title{ 
Polynomial-Time Algorithms for Prime Factorization 
and Discrete Logarithms on a Quantum Computer\thanks{A preliminary version 
of this paper appeared
in the Proceedings of the 35th Annual Symposium on Foundations of Computer 
Science, Santa Fe, NM, Nov.\ 20--22, 1994, IEEE Computer Society Press, 
pp.\ 124--134.}
}
\author{Peter W. Shor\thanks{AT\&T Research,
Room 2D-149, 600 Mountain Ave., Murray Hill, NJ 07974.}
}
\date{}
\maketitle

\begin{abstract}
{
A digital computer is generally believed to be an efficient universal 
computing device; that is, it is believed able to simulate any 
physical computing device with an increase in computation time by
at most a polynomial factor.
This may not be true when quantum mechanics is taken into consideration.  
This paper considers
factoring integers and finding discrete logarithms, two problems 
which are generally thought to be hard on a classical computer and 
which have been used as the basis of several proposed cryptosystems.  
Efficient randomized algorithms are given for these two problems 
on a hypothetical quantum computer.  These algorithms take a number of 
steps polynomial in the input size, e.g., the number of digits of the 
integer to be factored.  
}
\end{abstract}

\vspace{\baselineskip}

{\bf Keywords:}
algorithmic number theory, prime factorization, discrete logarithms, 
Church's thesis, quantum computers, foundations of quantum mechanics, 
spin systems, Fourier transforms

\vspace{\baselineskip}
{\bf AMS subject classifications:}
81P10, 11Y05, 68Q10, 03D10
\pagestyle{myheadings}
\thispagestyle{plain}
\markboth{P.~W. SHOR}{FACTORING WITH A QUANTUM COMPUTER}

\sloppy
\raggedbottom

\vfill
\newpage

\section{Introduction}
One of the first results in the mathematics of computation, which underlies 
the subsequent development of much of theoretical computer science, 
was the distinction between computable and non-computable functions
shown in papers of Church [1936], Turing [1936], and Post [1936].
Central to this result is Church's thesis, 
which says that all computing devices can be simulated by a 
Turing machine.  This thesis greatly simplifies the study of
computation, since it reduces the potential field of study from
any of an infinite number of potential computing devices to 
Turing machines.  Church's thesis is not a 
mathematical theorem; to make it one would require a
precise mathematical description of a computing device.
Such a description, however, would leave open the possibility of 
some practical computing
device which did not satisfy this precise mathematical description, and thus
would make the resulting
mathematical theorem weaker than Church's original thesis.

With the development of practical computers, it has become apparent
that the distinction between computable and non-computable functions 
is much too coarse; computer scientists are now interested in 
the exact efficiency with which specific functions can be computed.  This
exact efficiency, on the other hand, is too precise a quantity to
work with easily.  The generally accepted
compromise between coarseness and precision distinguishes 
efficiently and inefficiently computable functions by whether 
the length of the computation scales polynomially or superpolynomially 
with the input size.  The class of problems which can be
solved by algorithms having a number of steps polynomial in the
input size is known as~P.  

For this classification to make sense, we need it to be machine-independent.
That is, we need to know that whether a function
is computable in polynomial time is independent of the kind of computing 
device used.  This corresponds to the following quantitative 
version of Church's thesis, which
Vergis et al.\ [1986] have called the ``Strong Church's Thesis'' and 
which makes up half of the ``Invariance Thesis'' of van Emde Boas [1990].

{\sc Thesis {\rm (Quantitative Church's thesis).}} 
{\it 
Any physical computing 
device can be simulated by a Turing machine in a number of steps 
polynomial in the resources used by the computing device.}

In statements of this thesis, the Turing machine is sometimes augmented with 
a random number generator, as it has not yet been determined whether there 
are pseudorandom number generators which can efficiently simulate truly
random number generators for all purposes.
Readers who are not comfortable with Turing machines may think instead
of digital computers having an amount of memory that grows linearly with 
the length of the computation, as these two classes of computing machines
can efficiently simulate each other.

There are two escape clauses in the above thesis.  One of these is
the word ``physical.''  Researchers
have produced machine models that violate the 
above quantitative Church's thesis, 
but most of these have been ruled out by some reason for why they are
not ``physical,'' that is, why they could not be built and made
to work.  The other escape clause in the above thesis
is the word ``resources,'' the meaning of which is not completely specified 
above.  There are generally two resources which limit the ability of digital
computers to solve large problems: time (computation steps)
and space (memory).  There are more resources pertinent to analog 
computation; some proposed analog machines that seem able
to solve NP-complete problems in polynomial time have required the 
machining of exponentially precise parts, or an exponential amount of energy.
(See Vergis et al.\ [1986] and Steiglitz [1988]; this issue is also
implicit in the papers of Canny and Reif [1987] and Choi et al.\ [1995]
on three-dimensional shortest paths.)

For quantum computation, in addition to space and time,
there is also a third potentially important resource, precision.  For a 
quantum computer to work, at least in any currently envisioned 
implementation, it must be able to make changes in the quantum states 
of objects (e.g., atoms, photons, or nuclear spins).  These
changes can clearly not be perfectly accurate, but must contain some
small amount of 
inherent imprecision.  If this imprecision is constant (i.e., it does
not depend on the size of the input), then it is not
known how to compute any functions in polynomial time on a
quantum computer that cannot also be 
computed in polynomial time on a classical computer with a random
number generator.  However,
if we let the precision grow polynomially in the input size (that is,
we let the number of {\em bits} of precision grow logarithmically in
the input size), we appear
to obtain a more powerful type of computer.  Allowing the same polynomial
growth in precision does not appear to confer extra computing power to
classical mechanics, although allowing exponential growth in precision does
\cite{HaSi,VeStDi}.   

As far as we know, what 
precision is possible in quantum state manipulation is dictated not by 
fundamental physical laws but by the properties of the materials and 
the architecture with which a quantum computer is built.  It is
currently not clear which architectures, if any, will give high precision,
and what this precision will be.  If the precision
of a quantum computer
is large enough to make it more powerful than a classical computer,
then in order to understand its potential 
it is important to think of precision as a resource that can vary.  
Treating the precision as a large constant (even though it is almost
certain to be constant for any given machine) would be comparable to 
treating a classical digital computer as a finite automaton --- since 
any given computer has a fixed amount of memory, this view 
is technically correct; however, it is not particularly useful.

Because of the remarkable effectiveness of our mathematical models of 
computation, computer scientists have tended to forget that computation
is dependent on the laws of physics.  This can be seen in the 
statement of the quantitative Church's thesis in van Emde Boas [1990], where 
the word ``physical'' in the above phrasing is replaced with the word 
``reasonable.'' It is difficult to imagine any definition of 
``reasonable'' in this context which does not mean ``physically realizable,''
i.e., that this computing machine could actually be built and would work.

Computer scientists have become convinced of the truth of the
quantitative Church's thesis through the failure of all proposed
counter-examples.  Most of these proposed counter-examples have 
been based on the laws of classical mechanics; however, the 
universe is in reality
quantum mechanical.  Quantum mechanical objects often behave 
quite differently from how our intuition, based on classical mechanics, tells
us they should.  It thus seems plausible that the natural 
computing power of classical mechanics corresponds to Turing 
machines,\footnote{I believe that this question has not yet been settled 
and is worthy of further investigation.  See
Vergis et al.\ [1986], Steiglitz [1988], and Rubel [1989].  In particular, 
turbulence seems a good
candidate for a counterexample to the quantitative Church's thesis 
because the non-trivial dynamics on many length scales may make it
difficult to simulate on a classical computer.} 
while the natural computing power of 
quantum mechanics might be greater.  

The first person to look at the interaction between computation and
quantum mechanics appears to have been Benioff [1980, 1982a, 1982b].  
Although he did not ask whether quantum mechanics conferred extra power to 
computation, he showed that reversible unitary evolution was sufficient
to realize the computational power of a Turing machine,
thus showing that quantum mechanics is at least as powerful computationally
as a classical computer.  This work was fundamental in making
later investigation of quantum computers possible.

Feynman [1982,1986] seems to have been the first to suggest that quantum 
mechanics might be more powerful computationally than a Turing machine.
He gave arguments as to why quantum mechanics might be
intrinsically expensive computationally to simulate on a classical
computer.  He also raised the possibility of using a 
computer based on quantum mechanical principles to avoid this problem, thus 
implicitly asking the converse question: by using quantum mechanics in a 
computer can you compute more efficiently than on a classical computer?
Deutsch [1985, 1989] was the first to ask this question explicitly.
In order to study this question,
he defined both quantum Turing machines and quantum circuits and 
investigated some of their properties.

The question of whether using quantum mechanics in a computer allows one to 
obtain more computational power was more recently addressed by 
Deutsch and Jozsa [1992] and Berthiaume and Brassard [1992a, 1992b]. 
These papers showed that there are problems which quantum computers
can quickly solve exactly, but that classical computers can only solve
quickly with high probability and the aid of a random number generator.
However, these papers did not show how to solve any problem 
in quantum polynomial time that was not already known to be solvable in 
polynomial time with the aid of a random number generator, 
allowing a small probability of error;
this is the characterization of the complexity class BPP, which 
is widely viewed as the class of efficiently solvable problems.

Further work on this problem was stimulated by Bernstein and Vazirani [1993].
One of the results contained in 
their paper was an oracle problem (that is, a problem involving a 
``black box'' subroutine that the computer is allowed to perform,
but for which no code is accessible) which can be done in polynomial 
time on a quantum Turing machine but which requires super-polynomial time 
on a classical computer.  This result was improved
by Simon [1994], who gave a much simpler construction of an
oracle problem which takes polynomial time on a quantum computer but
requires {\em exponential} time on a classical computer.  Indeed, 
while Bernstein and Vaziarni's problem appears contrived, Simon's
problem looks quite natural.
Simon's algorithm inspired the work presented in this paper.

Two number theory problems which have been studied extensively but 
for which no polynomial-time algorithms have yet been discovered are
finding discrete logarithms and factoring integers
[Pomerance 1987, Gordon 1993, Lenstra and Lenstra 1993, Adleman and
McCurley 1995].  These problems
are so widely believed to be hard that several cryptosystems based on their
difficulty have been proposed, including the widely used
RSA public key cryptosystem developed by Rivest, Shamir, and Adleman [1978].
We show that these 
problems can be solved in polynomial time on a quantum computer
with a small probability of error.

Currently, nobody knows how to build a quantum computer, although it seems as
though it might be possible within the laws of quantum mechanics.  
Some suggestions have been made as to possible designs for such computers
[Teich et al.\ 1988, Lloyd 1993, 1994, Cirac and Zoller 1995,
DiVincenzo 1995, Sleator and Weinfurter 1995, Barenco et al.\ 1995b,
Chuang and Yamomoto 1995], but there will 
be substantial difficulty in building any of these 
\cite{Land1,Land2,Unru,ChLaShZu,PaSuEk}.  The most difficult 
obstacles appear to involve the decoherence of quantum superpositions 
through the interaction of the computer with the environment, and 
the implementation of quantum state transformations with enough precision
to give accurate results after many computation steps.
Both of these obstacles become more difficult as the size of the
computer grows, so it may turn out to be possible to
build small quantum computers, while scaling up to machines large enough 
to do interesting computations may present fundamental difficulties.

Even if no useful quantum computer is ever built, this research does 
illuminate the problem of simulating quantum mechanics on a classical 
computer.  Any method of doing this for an arbitrary Hamiltonian would 
necessarily be able to simulate a quantum computer.
Thus, any general method for simulating quantum mechanics with at most
a polynomial slowdown would lead to a polynomial-time algorithm for 
factoring.

The rest of this paper is organized as follows.  In \S2,
we introduce the model of quantum computation, the {\em quantum gate array,}
that we use in the rest of the paper.  In \S\S3 and~4,
we explain two subroutines that are used in our algorithms: 
reversible modular exponentiation in~\S3 
and quantum Fourier transforms in~\S4.  In \S5, we give our 
algorithm for prime factorization, and
in \S6, we give our algorithm for extracting discrete
logarithms.  In \S7, we give a brief discussion of the
practicality of quantum computation and suggest possible directions for
further work.

\section{Quantum computation}
In this section we give a brief introduction to quantum computation,
emphasizing the properties that we will use.  
We will describe only {\em quantum gate arrays,} or {\em quantum
acyclic circuits,} which are analogous to acyclic circuits in
classical computer science.  For other models 
of quantum computers, see references
on quantum Turing machines \cite{Deut89,BeVa,Yao} and quantum cellular 
automata [Feynman 1986, Margolus 1986, 1990, Lloyd 1993, Biafore 1994].
If they are allowed a small probability of 
error, quantum Turing machines and quantum gate arrays can compute the same 
functions in polynomial time \cite{Yao}.  This may also be true for the
various models
of quantum cellular automata, but it has not yet been proved.
This gives evidence that the class of functions computable in quantum
polynomial time with a small probability of error is robust, in that it
does not depend on the exact architecture of a quantum computer.  By analogy
with the classical class BPP, this class is called BQP.

Consider a system with $n$ components, each of which can have two states.
Whereas in classical physics, a complete description of the
state of this system requires
only $n$ bits, in quantum physics, a complete description of the state of 
this system requires $2^n-1$ complex numbers.  To be more
precise, the state of the quantum
system is a point in a $2^n$-dimensional vector space.
For each of the $2^n$ possible classical
positions of the components, there is a basis state 
of this vector space which we represent, for example,
by $\|0 1 1 \cdots 0\>$ meaning that the first bit is~0, the second
bit is~1, and so on.  Here, the {\em ket} notation $\|x\>$ means that
$x$ is a (pure) quantum state.  (Mixed states will not be discussed in
this paper, and thus we do not define them; see a quantum theory book
such as Peres [1993] for this definition.)
The {\em Hilbert space} associated with this quantum system is
the complex vector space with these $2^n$ states as basis vectors, and
the state of the system at any time is represented by a unit-length vector
in this Hilbert space.
As multiplying this state vector by a unit-length complex phase does 
not change any behavior of the state, we need only $2^n-1$ complex 
numbers to completely describe the state.
We represent this superposition of states as
\begin{equation}
\sum_{i=0}^{2^n-1} a_i \| S_i \>,
\end{equation}
where the amplitudes $a_i$ are complex numbers such that 
$\sum_i \left| a_i \right| ^2 =1$ and 
each $\| S_i \>$ is a basis vector of the Hilbert space. 
If the machine is measured (with respect to this basis)
at any particular step, the probability of 
seeing basis state $\| S_i \>$ is $\left| a_i \right| ^2$; however,
measuring the state of the machine projects this state to the 
observed basis vector $\|S_i\>$.
Thus, looking at the machine during the computation will invalidate the rest 
of the computation.  In this paper, we only consider measurements with
respect to the canonical basis.  This does not greatly restrict our
model of computation, since measurements in other reasonable
bases could be simulated by first using quantum computation to
perform a change of basis and then performing a measurement in the
canonical basis.

In order to use a physical
system for computation, we must be able to change the state of the system.
The laws of quantum mechanics permit only unitary transformations of 
state vectors.  A unitary matrix is one whose conjugate transpose is 
equal to its inverse, and requiring state transformations to be represented 
by unitary matrices ensures that summing the probabilities
of obtaining every possible outcome will result in~1.
The definition of quantum circuits (and quantum Turing machines) only
allows {\em local} unitary transformations; that is, unitary transformations
on a fixed number of bits.  
This is physically justified because, 
given a general unitary transformation on $n$ bits, it is
not at all clear how one would efficiently implement it physically, whereas 
two-bit transformations can at least in theory be implemented by relatively 
simple physical systems \cite{CiZo,DiVi,SlWe,ChYa}.  While general $n$-bit
transformations can always be built out of two-bit transformations
[DiVincenzo 1995, Sleator and Weinfurter 1995, Lloyd 1995, Deutsch et al.\ 
1995], the number required will often be 
exponential in $n$ \cite{nine}.   Thus, the set of two-bit transformations 
form a set of building blocks for quantum circuits in a manner analogous 
to the way a universal set of classical gates (such as the
AND, OR and NOT gates) form a set of building blocks for classical circuits. 
In fact, for a universal set of quantum gates, it is sufficient to take
all one-bit gates and a single type of two-bit gate, the controlled NOT,
which negates the second bit if and only if the first bit is~1.

Perhaps an example will be informative at this point.  A quantum gate
can be expressed as a truth table: for each input basis vector we need
to give the output of the gate.  One such gate is:\\
\noindent
\begin{minipage}{\textwidth} 
\begin{eqnarray}
\|00\> &\rightarrow& \|00\> \nonumber \\
\|01\> &\rightarrow& \|01\> \label{exampletrtable} \\
\|10\> &\rightarrow& 
{\textstyle \frac{1}{\sqrt{2}}} (\|10\> + \|11\>) \nonumber \\
\|11\> &\rightarrow& 
{\textstyle \frac{1}{\sqrt{2}}} (\|10\> - \|11\>). \nonumber
\end{eqnarray}
\end{minipage}

\vspace{\baselineskip}

\noindent
Not all truth tables correspond to physically feasible quantum gates,
as many truth tables will not give rise to unitary transformations.

The same gate can also be represented as a matrix. The rows correspond
to input basis vectors.  The columns correspond to output basis vectors.
The $(i,j)$ entry gives, when the $i$th basis vector is input to the gate, 
the coefficient of the $j$th basis vector in the corresponding output of 
the gate.  The truth table above would then correspond to the following 
matrix:
\begin{equation}
\begin{array}{c|cccc|l}
\multicolumn{1}{c}{} & \|00\> & \|01\> & \|10\> 
& \multicolumn{1}{c}{\|11\>} & \\*[.5ex]
\|00\> \ \  & 1 & 0 & 0 & 0 & \\*[.5ex]
\|01\> \ \  & 0 & 1 & 0 & 0 & \\*[.5ex]
\|10\> \ \  & 0 & 0 & \frac{1}{\sqrt{2}}  & \frac{1}{\sqrt{2}} & \\*[.5ex]
\|11\> \ \  & 0 & 0 & \frac{1}{\sqrt{2}} & -\frac{1}{\sqrt{2}}\phantom{-} &.
\end{array}
\label{examplematrix}
\end{equation}
A quantum gate is feasible if and only if the corresponding matrix is 
unitary, i.e., its inverse is its conjugate transpose.

Suppose our
machine is in the superposition of states
\begin{equation}
\textstyle
\frac{1}{\sqrt{2}} \|10\> - \frac{1}{\sqrt{2}} \|11\> 
\label{examplevector}
\end{equation}
and we apply the unitary transformation represented
by (\ref{exampletrtable}) and
(\ref{examplematrix}) to this state.  The resulting output will be the
result of multiplying the vector (\ref{examplevector}) by 
the matrix (\ref{examplematrix}).  The machine will thus go to the
superposition of states
\begin{equation}
\textstyle
\frac{1}{2}\left(\|10\> + \|11\>\right) 
-
\frac{1}{2}\left(\|10\> - \|11\>\right) \;=\;
\|11\>.
\end{equation}
This example shows the potential effects of interference on quantum
computation.  Had we started with either the state $\|10\>$ or the
state $\|11\>$, there would have been a chance of observing the
state $\|10\>$ after the application of the gate (\ref{examplematrix}).  
However, when we start with a superposition of these two states, the
probability amplitudes for the state $\|10\>$ cancel, and we have no 
possibility of observing $\|10\>$ after the application of the gate.
Notice that the output of the
gate would have been $\|10\>$ instead of $\|11\>$ had 
we started with the superposition of states
\begin{equation}
\textstyle
\frac{1}{\sqrt{2}} \|10\> + \frac{1}{\sqrt{2}} \|11\> 
\end{equation}
which has the same probabilities of being in any particular configuration
if it is observed as does the superposition (\ref{examplevector}).

If we apply a gate to only two bits of a longer basis vector (now
our circuit must have more than two wires), we multiply the
gate matrix by
the two bits to which the gate is applied, and leave the other bits alone.
This corresponds to multiplying the whole state by the tensor product
of the gate matrix on those two bits with the 
identity matrix on the remaining bits.

A quantum gate array is a set of quantum gates with 
logical ``wires'' connecting
their inputs and outputs.  The input to the 
gate array, possibly along with extra work bits that
are initially set to~0, is fed through a sequence of quantum gates.
The values of the bits are observed after the last quantum gate,
and these values are the output.  To compare
gate arrays with quantum Turing machines, we need to add conditions
that make gate arrays a {\em uniform} complexity class.  In other words,
because there is a different gate array for each size of input, we
need to keep the designer of the gate arrays from hiding non-computable
(or hard to compute) information in the arrangement of the gates.  
To make quantum gate arrays uniform, we must add two things to the
definition of gate arrays.  The first is the standard requirement that
the design of the gate array be produced by a polynomial-time (classical)
computation.  The second requirement 
should be a standard part of the definition of analog complexity classes, 
although since analog complexity classes have not
been widely studied, this requirement is much less widely known.
This requirement is that the entries in the unitary matrices describing
the gates must be computable numbers.  Specifically, the first $\log n$ 
bits of each entry should be classically computable in time polynomial in~$n$
\cite{Solo}.
This keeps non-computable (or hard to compute) information 
from being hidden in the bits of the amplitudes of the quantum gates.

\section{Reversible logic and modular exponentiation}
The definition of quantum gate arrays gives rise to completely reversible
computation.  That is, knowing the quantum state on the wires leading out
of a gate tells uniquely what the quantum state must have been on the
wires leading into that gate.  This is a reflection of the fact that,
despite the macroscopic arrow of time, the laws of physics appear to
be completely reversible.   This would seem to imply that anything built 
with the laws
of physics must be completely reversible; however, classical computers 
get around this fact by dissipating energy and thus making their 
computations thermodynamically irreversible.  This appears
impossible to do for quantum computers because superpositions of 
quantum states need to be maintained throughout the computation.  Thus, 
quantum computers necessarily have to use reversible computation.  
This imposes extra costs when doing classical computations on a quantum 
computer, as is sometimes necessary in subroutines of quantum
computations.

Because of the reversibility of quantum computation,
a deterministic computation is performable on a quantum
computer only if it is reversible.  
Luckily, it has already been shown that any deterministic computation
can be made reversible \cite{Lece,Benn73}.  In fact, reversible
classical gate arrays have been studied.  Much like the result that any 
classical computation can be done using NAND gates, there are also universal
gates for reversible computation.  Two of these are Toffoli gates 
\cite{Toff} and Fredkin gates \cite{FrTo}; these are illustrated in
Table~\ref{revgates}.  

\begin{table}
\caption{Truth tables for Toffoli and Fredkin gates.}
\footnotesize
\hspace*{\fill}
\begin{tabular}%
{|@{\hspace{1em}}ccc@{\hspace{1em}}|@{\hspace{1em}}ccc@{\hspace{1em}}|}
\multicolumn{6}{c}{Toffoli Gate}\\
\multicolumn{6}{c}{}\\
\hline
\multicolumn{3}{|c|@{\hspace{1em}}}{INPUT}&
\multicolumn{3}{@{\hspace{-1em}}c@{\hspace{0em}}|}{OUTPUT}\\
\hline
0 & 0 & 0 & 0 & 0 & 0 \\
0 & 0 & 1 & 0 & 0 & 1 \\
0 & 1 & 0 & 0 & 1 & 0 \\
0 & 1 & 1 & 0 & 1 & 1 \\
1 & 0 & 0 & 1 & 0 & 0 \\
1 & 0 & 1 & 1 & 0 & 1 \\
1 & 1 & 0 & 1 & 1 & 1 \\
1 & 1 & 1 & 1 & 1 & 0 \\
\hline
\end{tabular}
\hspace*{\fill}
\begin{tabular}%
{|@{\hspace{1em}}ccc@{\hspace{1em}}|@{\hspace{1em}}ccc@{\hspace{1em}}|}
\multicolumn{6}{c}{Fredkin Gate}\\
\multicolumn{6}{c}{}\\
\hline
\multicolumn{3}{|c|@{\hspace{1em}}}{INPUT}&
\multicolumn{3}{@{\hspace{-1em}}c@{\hspace{0em}}|}{OUTPUT}\\
\hline
0 & 0 & 0 & 0 & 0 & 0 \\
0 & 0 & 1 & 0 & 1 & 0 \\
0 & 1 & 0 & 0 & 0 & 1 \\
0 & 1 & 1 & 0 & 1 & 1 \\
1 & 0 & 0 & 1 & 0 & 0 \\
1 & 0 & 1 & 1 & 0 & 1 \\
1 & 1 & 0 & 1 & 1 & 0 \\
1 & 1 & 1 & 1 & 1 & 1 \\
\hline
\end{tabular}
\hspace*{\fill}
\label{revgates}
\end{table}

The Toffoli gate is just a controlled controlled NOT, i.e., the
last bit is negated if and only if the first two bits are~1.
In a Toffoli gate, if the third input bit is set to~1, then the
third output bit is the NAND of the first two input bits.  Since
NAND is a universal gate for classical gate arrays, this shows
that the Toffoli gate is universal.  In a Fredkin gate, the
last two bits are swapped if the first bit is~0, and left untouched
if the first bit is~1.  For a Fredkin gate, if the third input bit 
is set to~0, the second output bit is the AND of the first two input
bits; and if the last two input bits are set to 0 and 1 respectively,
the second output bit is the NOT of the first input bit.  Thus, both
AND and NOT gates are realizable using Fredkin gates, showing that
the Fredkin gate is universal.

 From results on 
reversible computation \cite{Lece,Benn73}, we can compute
any polynomial time function $F(x)$ as long as we keep the input $x$ 
in the computer.  We do this by adapting the method for computing the
function~$F$ non-reversibly.  These results can easily be extended
to work for gate arrays \cite{Toff,FrTo}.
When AND, OR or NOT gates are changed to Fredkin or Toffoli gates, 
one obtains both additional input bits, which must be preset to specified
values, and additional output bits, which contain the information needed to 
reverse the computation.  While the additional input bits 
do not present difficulties in designing quantum computers, the
additional output bits do, because unless they are all reset to~0,
they will affect the interference patterns in quantum computation.
Bennett's method for resetting these bits to 0 is shown in the top half
of Table~\ref{reverse}.  A non-reversible gate array may thus be turned 
into a reversible gate array as follows.  First, duplicate the 
input bits as many times as necessary (since each input bit could be used
more than once by the gate array). Next, keeping one copy of the input
around, use Toffoli and Fredkin gates to simulate non-reversible gates,
putting the extra output bits into the RECORD register. These extra output
bits preserve enough of a record of the operations to enable the computation
of the gate array to be reversed. 
Once the output $F(x)$ has been computed, copy it into a register that
has been preset to zero, and then undo the computation to erase both
the first OUTPUT register and the RECORD register.  

\begin{table}[ht]
\caption{Bennett's method for making a computation reversible.}
\footnotesize
\begin{center}
\begin{tabular}
{|@{\hspace{2em}}p{7.5em}p{7.5em}p{7.5em}p{7.5em}@{\hspace{-1em}}|}
\hline
INPUT & - - - - - - & - - - - - - & - - - - - - \\
INPUT & OUTPUT & RECORD($F$) & - - - - - - \\
INPUT & OUTPUT & RECORD($F$) & OUTPUT \\
INPUT & - - - - - - & - - - - - - & OUTPUT \\
\hline
INPUT & INPUT & RECORD($F^{-1}$) & OUTPUT \\
- - - - - - & INPUT & RECORD($F^{-1}$) & OUTPUT \\
- - - - - - & - - - - - - & - - - - - - & OUTPUT\\
\hline
\end{tabular}
\hspace{3em}
\end{center}
\label{reverse}
\end{table}

To erase $x$ and replace it with $F(x)$, in addition to a
polynomial-time algorithm for~$F$, we also need
a polynomial-time algorithm for computing $x$ 
from $F(x)$; i.e., we need that $F$ is one-to-one and
that both $F$ and $F^{-1}$ are polynomial-time computable.  
The method for this computation
is given in the whole of Table~\ref{reverse}.
There are two stages to this computation.  The first is the same as
before, taking $x$ to $(x,F(x))$.  For the second stage, shown in the
bottom half of Table~\ref{reverse}, note that if
we have a method to compute $F^{-1}$ non-reversibly in polynomial time, 
we can use the same technique to reversibly
map $F(x)$ to $(F(x), F^{-1}(F(x))) = (F(x),x)$.  However, since this
is a reversible computation, we can reverse it
to go from $(x,F(x))$ to $F(x)$.  Put together,
these two pieces take $x$ to $F(x)$.

The above discussion shows that computations can be made reversible for
only a constant factor cost in time, but the above method uses as much space
as it does time.  If the classical computation requires much less space 
than time, then making it reversible in this manner will result
in a large increase 
in the space required.  There are methods that do not use as much
space, but use more time, to 
make computations reversible \cite{Benn89,LeSh}.  While there is no 
general method that does not cause an increase in either space or time, 
specific algorithms can sometimes be made reversible without paying 
a large penalty in either space or time; at the end of this
section we will show how to do 
this for modular exponentiation, which is a subroutine necessary for 
quantum factoring.

The bottleneck in the quantum factoring algorithm; i.e., the piece
of the factoring algorithm that consumes the most time and
space, is modular exponentiation.  The modular exponentiation 
problem is, given $n$, $x$, and $r$, find $x^r \mod{n}$.  
The best classical method for doing this is to repeatedly square of 
$x\mod{n}$ to get $x^{2^i}\mod{n}$ for $i\leq\log_2 r$, and then multiply a 
subset of these powers $\bmod{n}$ to get $x^r\mod{n}$.  If
we are working with $l$-bit numbers, this requires $O(l)$ squarings
and multiplications of $l$-bit numbers $\bmod{n}$.  Asymptotically, 
the best classical result for gate arrays for 
multiplication is the Sch\"onhage--Strassen algorithm \cite{ScSt,Knut,Scho}.
This gives a gate array for integer multiplication that uses
$O(l \log l \log \log l)$ gates to multiply two $l$-bit numbers.  Thus,
asymptotically, modular exponentiation requires $O(l^2 \log l \log \log l)$
time.  Making this reversible would na\"\i vely cost the same amount in 
space; however, one can reuse the space used in the repeated squaring part 
of the algorithm, and thus reduce the amount of space needed to essentially
that required for multiplying two $l$-bit numbers; one simple
method for reducing this space (although 
not the most versatile one) will be given later in this 
section.  Thus, modular exponentiation can be done in 
$O(l^2 \log l \log \log l)$ time and $O(l \log l \log \log l )$ space.  

While the Sch\"onhage--Strassen algorithm is the best 
multiplication algorithm discovered
to date for large~$l$, it does not scale well for small~$l$.
For small numbers, the best gate arrays for multiplication essentially use
elementary-school longhand multiplication in binary.  This method requires
$O(l^2)$ time to multiply two $l$-bit numbers, and thus modular 
exponentiation requires $O(l^3)$ time with this method.  These gate arrays 
can be made reversible, however, using only $O(l)$ space.  

We will now give the method for constructing a reversible gate array that 
takes only $O(l)$ space and $O(l^3)$ time to
compute $(a,x^a\mod{n})$ from~$a$, where $a$, $x$, and $n$
are $l$-bit numbers.  The basic building
block used is a gate array that takes $b$ as input and outputs 
$b+c\mod{n}$.  Note that here $b$ is the gate array's input but $c$ and $n$
are built into the structure of the gate array.
Since addition $\bmod{n}$ is
computable in $O(\log n)$ time classically, this reversible gate array 
can be made with only $O(\log n)$ gates and $O(\log n)$ work
bits using the techniques explained earlier in this section.

The technique we use for computing $x^a\mod{n}$ is essentially the same
as the classical method.  First, by repeated squaring we compute
$x^{2^i}\mod{n}$ for all $i < l$.  Then, to obtain $x^a\mod{n}$ we 
multiply the powers $x^{2^i}$ $\bmod{n}$
where $2^i$ appears in the binary expansion of~$a$.
In our algorithm for factoring~$n$, we only
need to compute $x^{a}\mod{n}$ where $a$ is in a superposition of states,
but $x$ is some fixed integer.  This makes things much easier,
because we can use a reversible gate array where $a$ is 
treated as input, but where $x$ and $n$ are built
into the structure of the gate array.  Thus, we can use the algorithm
described by the following pseudocode;
here, $a_i$ represents the $i$th bit of $a$ in binary, 
where the bits are indexed from right to left and the rightmost
bit of $a$ is~$a_0$.

\vspace{\baselineskip} 

\noindent
\begin{minipage}{\textwidth} 
{\tt
\begin{tabbing}
\ \ \ \ \ \ \ \ \ \= \ \ \ \ \= \ \ \ \ \= \kill
\tab {\it power} := 1 \\
\tab for {\it i} = 0 to {\it l}$\,-$1 \\
\tab \tab if ( {\it a}$_i$ $==$ 1 ) then \\
\tab \tab \tab {\it power} := {\it power} $*$ {\it x}$^{\,2^i}$ $\mod{n}$ \\
\tab \tab endif\\
\tab endfor
\end{tabbing}
}
\end{minipage}

\vspace{\baselineskip}

\noindent
The variable $a$ is left unchanged by the code and $x^a\mod{n}$ 
is output as the variable ${\it power}$.  Thus, this code takes
the pair of values $(a,1)$ to $(a,x^a\mod{n})$.

This pseudocode can easily be turned into a gate array;
the only hard part of this is the fourth line, where we multiply the
variable {\it power} by $x^{2^i}$ $\bmod{n}$; to do this we need to use a 
fairly complicated gate array as a subroutine.  
Recall that $x^{2^i}\mod{n}$ can be computed classically and then 
built into the 
structure of the gate array.  Thus, to implement this line, we need
a reversible gate array that takes $b$ as input and gives $bc \mod{n}$ as
output, where the structure of the gate array can depend on $c$ and~$n$.  
Of course, this step
can only be reversible if $\gcd(c,n) =1$, i.e., if
$c$ and $n$ have
no common factors, as otherwise two distinct values of $b$ will be 
mapped to the same value of $bc \mod{n}$; this case is 
fortunately all we need
for the factoring algorithm.  We will show how
to build this gate array in two stages.  The first stage is directly 
analogous to exponentiation by repeated multiplication; we obtain 
multiplication from repeated addition $\bmod{n}$.  Pseudocode for this 
stage is as follows.  

\vspace{\baselineskip}

\noindent
\begin{minipage}{\textwidth} 
{\tt
\begin{tabbing}
\ \ \ \ \ \ \ \ \ \= \ \ \ \ \= \ \ \ \ \= \kill
\tab {\it result} := 0\\
\tab for {\it i} = 0 to {\it l}$\,-$1\\
\tab \tab if  ( {\it b}$_i$ $==$ 1 ) then \\
\tab \tab \tab {\it result} := {\it result} $+$ 2$^{i}${\it c}  $\mod{n}$\\
\tab \tab endif\\
\tab endfor
\end{tabbing}
}
\end{minipage}

\vspace{\baselineskip}

\noindent
Again, $2^ic\mod{n}$ can be precomputed and built into the 
structure of the gate array.

The above pseudocode takes $b$ as input, and gives $(b,bc\mod{n})$ as 
output.  To get the desired result, we now
need to erase~$b$.  Recall that $\gcd(c,n)=1$, so
there is a $c^{-1}\mod{n}$ with $c \, c^{-1} \equiv 1\mod{n}$.  
Multiplication by this $c^{-1}$ could be used to reversibly take 
$bc\mod{n}$ to $(bc\mod{n},bcc^{-1}\mod{n}) = (bc\mod{n},b)$.   This
is just the reverse of the operation we want, and since we are working
with reversible computing, we can turn this operation around to erase~$b$.  
The pseudocode for this follows.

\vspace{\baselineskip}

\noindent
\begin{minipage}{\textwidth} 
{\tt
\begin{tabbing}
\ \ \ \ \ \ \ \ \ \= \ \ \ \ \= \ \ \ \ \= \kill
\tab for {\it i} = 0 to {\it l}$\,-$1\\
\tab \tab if  ( {\it result}$_{\,i}$ $==$ 1 ) then\\
\tab \tab \tab {\it b} := {\it b} $-$ 2$^{i}${\it c}$^{-1}$ $\mod{n}$\\
\tab \tab endif\\
\tab endfor
\end{tabbing}
}
\end{minipage}

\vspace{\baselineskip}

\noindent
As before, {\it result}$_{\,i}$ is the $i$th bit of {\it result}.

Note that at this stage of the computation, $b$ should be~0.  
However, we did not set
$b$ directly to zero, as this would not have been a reversible
operation and thus impossible on a quantum computer,
but instead we did a relatively complicated sequence
of operations which ended with $b=0$ and which in fact depended on 
multiplication being a group $\bmod{n}$.  At
this point, then, we could do something somewhat sneaky: we could
measure $b$ to see if it actually is~0.  If it is not, we know that
there has been an error somewhere in the quantum computation, 
i.e., that the results are worthless and we should stop the computer 
and start over again.  However, if we do find that $b$ is~0, then we know 
(because we just observed it) that it is now exactly~0.  This 
measurement thus may bring the quantum computation back on track 
in that any amplitude that $b$ had for being non-zero has been
eliminated.  Further, because the probability that we observe a state 
is proportional to the square of the amplitude of that state, depending
on the error model, doing the 
modular exponentiation and measuring $b$ every time that we know 
that it should be 0 may have a higher probability of overall success
than the same computation done without the repeated measurements of~$b$; 
this is the {\em quantum watchdog} (or {\em quantum Zeno}) effect 
\cite{Pere2}.  
The argument above does not actually show that repeated measurement of
$b$ is indeed beneficial, because there is a cost (in time, if nothing
else) of measuring~$b$.  Before this is implemented, then, it should 
be checked with analysis or experiment that the benefit of 
such measurements exceeds their cost.  However, I believe that 
partial measurements such as this one are a promising way of trying 
to stabilize quantum computations.

Currently, Sch\"onhage--Strassen is the algorithm of choice for 
multiplying very large numbers, and longhand multiplication is the algorithm 
of choice for small numbers.
There are also multiplication algorithms which have efficiencies between
these two algorithms, and which are the best
algorithms to use for intermediate length numbers \cite{KaOf,Knut,ScGrVe}.  
It is not clear which algorithms are best for which size numbers.  
While this may be known to some extent for classical computation 
\cite{ScGrVe}, using data on which algorithms work better on classical 
computers could be misleading for two reasons:  First, classical computers 
need not be reversible, and the cost of making an algorithm reversible 
depends on the algorithm.  Second, existing computers generally have 
multiplication for 32- or 64-bit numbers built into their hardware, 
and this will increase the optimal changeover points to asymptotically
faster algorithms; further, some multiplication algorithms can take 
better advantage of this hardwired multiplication than others.  Thus, 
in order to program quantum computers most efficiently, work needs
to be done on the best way of implementing elementary arithmetic 
operations on quantum computers.  One tantalizing fact is that the
Sch\"onhage--Strassen fast multiplication algorithm uses the fast
Fourier transform, which is also the basis for all the fast algorithms
on quantum computers discovered to date; it is tempting to speculate
that integer multiplication itself might be speeded up by a quantum
algorithm; if possible, this would result in a somewhat faster asymptotic
bound for factoring on a quantum computer, and indeed could even
make breaking RSA on a quantum computer asymptotically faster 
than encrypting with RSA on a classical computer.

\section{Quantum Fourier transforms}
Since quantum computation deals with unitary transformations, it is helpful
to be able to build certain useful unitary transformations.  In
this section we give a technique for constructing in polynomial time
on quantum computers one particular unitary transformation,
which is essentially a discrete Fourier transform.  This transformation
will be given as a matrix, with both rows and columns indexed
by states.  These states correspond to binary
representations of integers on the computer; in particular, the rows 
and columns will be indexed beginning with 0 unless otherwise specified.

This transformations is as follows.
Consider a number $a$ with $0 \leq a < q$ for some
$q$ where the number of bits of $q$ is polynomial.  We will perform the
transformation that takes the state $\|a\>$ to the state
\begin{equation}
\frac{1}{q^{1/2}} \sum_{c=0}^{q-1} \|c\> \exp(2 \pi i ac/q).
\label{ft}
\end{equation}
That is, we apply
the unitary matrix whose 
$(a,c)$ entry is
$\frac{1}{q^{1/2}}\exp(2 \pi i ac/q)$.   
This Fourier transform is at the heart of our algorithms, and we call
this matrix~$A_q$.  

Since we will use $A_q$ for $q$ of exponential size, we must show how this 
transformation can be done in polynomial time.  In this paper, we will
give a simple construction for $A_q$ when $q$ is a power of~2 that was
discovered independently by Coppersmith [1994] and Deutsch 
[see Ekert and Jozsa 1995].  This construction is essentially the
standard fast Fourier transform (FFT) algorithm \cite{Knut}
adapted for a quantum computer; the following description of it 
follows that of Ekert and Jozsa [1995]. 
In the earlier version of this paper [Shor 1994], we gave a 
construction for $A_q$ when $q$ was in the special class of smooth numbers 
with small prime power factors.  In fact, Cleve [1994] has shown how 
to construct $A_q$ for all smooth numbers $q$
whose prime factors are at most $O(\log n)$.  

Take $q=2^l$, and let us represent an integer $a$ in binary as
$\|a_{l-1}a_{l-2}\ldots a_0\>$.
For the quantum Fourier transform~$A_q$, we only need to use two types of 
quantum gates.  These gates are $R_j$, which operates on the $j$th
bit of the quantum computer:
\begin{equation}
R_j \ = \ 
\begin{array}{c|cc|l}
\multicolumn{1}{c}{} & \|0\> & \multicolumn{1}{c}{\|1\>} & \\*[.5ex]
\|0\> \ \  & \frac{1}{\sqrt{2}} & \phantom{-} \frac{1}{\sqrt{2}} & \\*[.5ex]
\|1\> \ \  & \frac{1}{\sqrt{2}}  & -\frac{1}{\sqrt{2}}& ,
\end{array}
\label{Rmatrix}
\end{equation}
and
$S_{j,k}$, which operates on the bits in positions $j$ and $k$ with $j < k$:
\begin{equation}
S_{j,k} \ = \ 
\begin{array}{c|cccc|l}
\multicolumn{1}{c}{} & \|00\> & \|01\> & \|10\> & 
                      \multicolumn{1}{c}{\|11\>} & \\*[.5ex]
\|00\> \ \  & 1 & 0 & 0 & 0 & \\*[.5ex]
\|01\> \ \  & 0 & 1 & 0 & 0 & \\*[.5ex]
\|10\> \ \  & 0 & 0 & 1 & 0 & \\*[.5ex]
\|11\> \ \  & 0 & 0 & 0 & e^{i\theta_{k-j}} & ,
\end{array}
\label{Smatrix}
\end{equation}
where $\theta_{k-j} = \pi/2^{k-j}$.
To perform a quantum Fourier transform, we apply the matrices in the
order (from left to right)
\begin{equation}
\, R_{l-1}\, S_{l-2,l-1}\, R_{l-2}\, S_{l-3,l-1}\, S_{l-3,l-2}\, R_{l-3} 
\ldots R_1 \,S_{0,l-1}\, S_{0,l-2} \ldots S_{0,2}\, S_{0,1}\, R_0\, ;
\end{equation}
that is, we apply the gates $R_j$ in reverse order from $R_{l-1}$ to~$R_0$,
and between $R_{j+1}$ and $R_j$ we apply all the gates $S_{j,k}$ where $k>j$.
For example, on 3 bits, the matrices would be applied in the order
$R_2 S_{1,2} R_1 S_{0,2} S_{0,1} R_0$.  
To take the Fourier transform
$A_q$ when $q=2^l$, we thus need to use $l(l-1)/2$ quantum gates.

Applying this sequence of transformations will result in a quantum state 
$\frac{1}{q^{1/2}} \sum_b \exp(2 \pi i ac /q) \|b\>$, 
where $b$ is the bit-reversal of~$c$,
i.e., the binary number obtained by reading the bits of $c$ from right 
to left.  Thus, to obtain the actual quantum Fourier transform, we need 
either to do further computation to
reverse the bits of $\|b\>$ to obtain $\|c\>$, or to leave these bits
in place and read them in reverse order; either alternative is easy
to implement. 

To show that this operation actually performs a quantum Fourier transform,
consider the amplitude of going from $\|a\> = \|a_{l-1} \ldots a_0\>$
to $\|b\> = \|b_{l-1} \ldots b_0\>$.  First, the factors of $1/\sqrt{2}$
in the $R$ matrices multiply to produce a factor of $1/q^{1/2}$ overall;
thus we need only worry about the $\exp(2 \pi i ac/q)$ phase 
factor in the
expression (\ref{ft}).  The matrices $S_{j,k}$ do not change the values of 
any bits, but merely change their phases.  There is thus only one way to 
switch the $j$th bit from $a_j$ to~$b_j$, and that is to use the appropriate
entry in the matrix~$R_j$.  This entry adds $\pi$ to the phase if the
bits $a_j$ and $b_j$ are both~1, and leaves it unchanged otherwise.
Further, the matrix $S_{j,k}$ adds $\pi/2^{k-j}$
to the phase if $a_j$ and $b_k$ are both 1 and leaves it unchanged otherwise.
Thus, the phase on the path from $\|a\>$ to $\|b\>$ is
\begin{equation}
\sum_{0 \leq j < l} \pi a_jb_j + 
\sum_{0\leq j<k < l} \frac{\pi}{2^{k-j}} a_j b_k.
\end{equation}
This expression can be rewritten as
\begin{equation}
 \sum_{0 \leq j \leq k < l} \frac{\pi}{2^{k-j}} a_j b_k.
\end{equation}
Since $c$ is the bit-reversal of~$b$, this expression can be 
further rewritten as
\begin{equation}
 \sum_{0 \leq j \leq k < l} \frac{\pi}{2^{k-j}} a_j c_{l-1-k}.
\end{equation}
Making the substitution $l-k-1$ for $k$ in this sum, we get 
\begin{equation}
 \sum_{0 \leq j+k < l} 2 \pi \frac{2^j 2^k}{2^l} a_j c_{k} 
\end{equation}
Now, since adding multiples of $2 \pi$ do not affect the phase, we obtain
the same phase if we sum over all $j$ and $k$ less than~$l$, obtaining
\begin{equation}
 \sum_{j,k =0}^{l-1} 2 \pi \frac{2^j 2^k}{2^l} a_j c_{k} = \frac{2\pi}{2^l}\;
\sum_{j=0}^{l-1} 2^j a_j \;\sum_{k=0}^{l-1} 2^k c_k,
\end{equation}
where the last equality follows
from the distributive law of multiplication.
Now, $q=2^l$, $a = \sum_{j=0}^{l-1}2^j a_j$, and similarly for~$c$, so 
the above expression is equal to
$2 \pi ac/q$,
which is the phase for the amplitude of $\|a\> \rightarrow \|c\>$ in
the transformation (\ref{ft}).

When $k-j$ is large in the gate $S_{j,k}$ in (\ref{Smatrix}), we are 
multiplying by a 
very small phase factor.  This would be very difficult to do
accurately physically, and thus it would be somewhat disturbing if this
were necessary for quantum computation.  Luckily, Coppersmith [1994]
has shown that one can define an approximate Fourier transform that ignores
these tiny phase factors, but which approximates the Fourier transform
closely enough that it can also be used for factoring.  In fact,
this technique reduces the number of quantum gates needed for the
(approximate) Fourier transform considerably, as it leaves out most of
the gates~$S_{j,k}$.

\section{Prime factorization}
It has been known since before Euclid that every integer $n$ is uniquely
decomposable into a product of primes.  Mathematicians have been 
interested in the question of how to factor a number into this 
product of primes for nearly as long.  It was only in the 1970's, 
however, that researchers applied the paradigms of theoretical computer
science to number theory, and looked at the asymptotic running times
of factoring algorithms \cite{Adle}.  This has resulted in a great 
improvement in the efficiency of factoring algorithms.   The best
factoring algorithm asymptotically is currently the number field sieve
\cite{LeLeMaPo,LeLe}, which in order to factor an integer $n$ takes 
asymptotic running time 
$\exp(c(\log n)^{1/3} (\log \log n)^{2/3})$ for some constant~$c$.  
Since the input, $n$, is only $\log n$ bits in length, this
algorithm is an exponential-time algorithm.
Our quantum factoring algorithm takes asymptotically
$O((\log n)^2 \linebreak[1] (\log \log n) \linebreak[1] (\log \log \log n))$ 
steps on a quantum computer, along with a polynomial (in $\log n$) amount 
of post-processing time on a classical computer that is
used to convert the output of the quantum computer to factors 
of~$n$.  While this post-processing could in principle be done on a quantum
computer, there is no reason not to use a classical computer if they are
more efficient in practice.

Instead of giving a quantum computer algorithm for factoring $n$ directly,
we give a quantum computer algorithm for finding the order of
an element $x$ in the multiplicative group
$\bmod{n}$; that is, the least integer $r$ such that 
$x^r \equiv 1 \mod{n}$.  It is known that using randomization,
factorization can be reduced to finding the order of an element \cite{Mill}; 
we now briefly give this reduction.

To find a factor of an odd number~$n$, given a method for computing
the order $r$ of~$x$, choose a random $x \mod{n}$,
find its order~$r$, and compute $\gcd(x^{r/2}-1,n)$.  Here,
$\gcd(a,b)$ is the greatest common divisor of $a$ and~$b$, i.e.,
the largest integer that divides both $a$ and~$b$.  The  
Euclidean algorithm \cite{Knut} can be used to compute $\gcd(a,b)$
in polynomial time.  Since
$(x^{r/2}-1)\linebreak[1](x^{r/2}+1) = x^{r}-1 \equiv 0 \mod{n}$, 
the $\gcd(x^{r/2}-1,n)$ fails to be
a non-trivial divisor of $n$ only if $r$ is odd or if $x^{r/2} 
\equiv -1 \mod{n}$.  Using this criterion, it can be shown that this 
procedure, when applied to a random~$x \mod{n}$,
yields a factor of $n$ with probability 
at least $1-1/2^{k-1}$, where $k$
is the number of distinct odd prime factors of~$n$.  A brief sketch of the
proof of this result follows.  Suppose that $n = \prod_{i=1}^{k} p_i^{a_i}$. 
Let $r_i$ be the order of $x \mod{p_i^{a_i}}$. 
Then $r$ is the least common multiple of all the~$r_i$.  
Consider the largest power of 2 dividing each~$r_i$.   The algorithm
only fails if all of these powers of 2 agree: if they are all~1, then
$r$ is odd and $r/2$ does not exist; if they are all equal and 
larger than~1, then
$x^{r/2} \equiv -1 \mod{n}$ since $x^{r/2} \equiv -1 \mod{p_i^{\alpha_i}}$
for every~$i$.  By the Chinese remainder
theorem [Knuth 1981, Hardy and Wright 1979, Theorem 121], choosing an 
$x \mod{n}$ at random is the same as choosing 
for each $i$ a number $x_i \mod{p_i^{a_i}}$ at random, where $p_i^{a_i}$ 
is the $i$th prime power
factor of~$n$.  The multiplicative group $\bmod{p^{\alpha}}$ for
any odd prime power $p^{\alpha}$ is
cyclic \cite{Knut}, so for any odd prime power $p_i^{a_i}$, the 
probability is at most $1/2$ of 
choosing an $x_i$ having any particular power of two as the largest divisor 
of its order~$r_i$.
Thus each of these powers of 2 has at most a 50\% probability
of agreeing with the previous ones, so all $k$ of them agree with
probability at most $1/2^{k-1}$,
and there is at least a $1-1/2^{k-1}$ chance that the $x$ we choose
is good.
This scheme will thus
work as long as $n$ is odd and not a prime power; finding factors of 
prime powers can be done efficiently with classical methods.

We now describe the algorithm for finding the 
order of $x \mod{n}$ on a quantum computer.
This algorithm will use two quantum registers which hold integers represented
in binary.  There will also be some amount of workspace.  This workspace
gets reset to 0 after each subroutine of our algorithm, so we will 
not include it when we write down the state of our machine. 

Given $x$ and~$n$, to find the order of~$x$, i.e., the least $r$ such 
that $x^r \equiv 1 \mod{n}$, we do the
following.  First, we find $q$, the power of 2 with $n^2 \leq q < 2 n^2$.  
We will not include $n$, $x$, or $q$ when we write down
the state of our machine, because we never change these values.
In a quantum gate array we need not even keep these values in
memory, as they can be built into the structure of the gate array.

Next, we put the first register
in the uniform superposition of states representing
numbers $a \mod{q}$.  This leaves our machine in state
\begin{equation}
\frac{1}{q^{1/2}} \sum_{a=0}^{q-1} \|a\>\|0\>.
\end{equation}
This step is relatively easy, since all it entails is putting each bit
in the first register into the superposition $\frac{1}{\sqrt{2}}(\|0\>
+\|1\>)$.

Next, we compute $x^a\mod{n}$ in the second register as described in~\S3.  
Since we keep $a$ in the first register
this can be done reversibly.  This leaves our machine in the state
\begin{equation}
\frac{1}{q^{1/2}} \sum_{a=0}^{q-1} \|a\>\|x^a\mod{n}\>.
\end{equation}

We then perform our Fourier transform $A_q$ on the first register,
as described in~\S4,
mapping $\|a\>$ to
\begin{equation}
\frac{1}{q^{1/2}} \sum_{c=0}^{q-1} \exp(2 \pi i ac/q) \|c\>.
\end{equation}
That is, we apply the unitary matrix
with the $(a,c)$ entry equal to $\frac{1}{q^{1/2}} \exp(2 \pi i ac/q)$.
 This leaves our machine in state
\begin{equation}
\frac{1}{q} \sum_{a=0}^{q-1} \sum_{c=0}^{q-1} 
\exp(2 \pi i ac / q) \|c\>\|x^a\mod{n}\>.
\end{equation}

Finally, we observe the machine.  It would be sufficient to observe solely
the value of $\|c\>$ in the first register, but for clarity we will 
assume that we observe both $\|c\>$
and $\|x^a\mod{n}\>$.
We now compute the probability that our machine ends in a particular
state $\|c,x^k\mod{n}\>$, where we may assume $0 \leq k < r$.
Summing over all possible ways to reach the state
$\|c,x^k\mod{n}\>$, we find that
this probability is
\begin{equation}
\left|
\frac{1}{q} \sum_{a:\, x^a\equiv x^k}
\exp(2 \pi i ac / q) 
\right|^2 .
\end{equation}
where the sum is over all $a$, $0\leq a < q$, such 
that $x^a \equiv x^k\mod{n}$.
Because the order of $x$ is~$r$, this sum is over 
all $a$ satisfying
$a \equiv k \mod{r}$.  
Writing $a=br+k$, we find that the above probability is
\begin{equation}
\left|
\frac{1}{q} \sum_{b=0}^{\left\lfloor (q-k-1)/r \right\rfloor} 
\exp(2 \pi i (br+k) c / q) 
\right|^2 .
\end{equation}
We can ignore the term of $\exp(2 \pi i kc/q)$, as it can
be factored out of the
sum and has magnitude~1.  We can also replace $rc$ with $\lr{rc}{q}$,
where $\lr{rc}{q}$ is the residue which is congruent to $rc\mod{q}$
and is in the range $-q/2 < \lr{rc}{q} \leq q/2$.  This leaves us with
the expression
\begin{equation}
\left|
\frac{1}{q} \sum_{b=0}^{\left\lfloor (q-k-1)/r \right\rfloor} 
\exp(2 \pi i b \lr{rc}{q} / q) 
\right|^2 .
\label{cprobs}
\end{equation}
We will now show that if $\lr{rc}{q}$
is small enough, all the amplitudes in this sum will be
in nearly the same direction (i.e., have close to the same phase), 
and thus make the sum large.
Turning the sum into an integral, we obtain
\begin{equation}
\frac{1}{q} \int_{0}^{\left\lfloor \frac{q-k-1}{r} \right\rfloor} 
\exp(2 \pi i b \lr{rc}{q} / q) db + 
O\left({\textstyle \frac{\lfloor (q-k-1)/r \rfloor}{q}}
\left( \exp(2\pi i \lr{rc}{q} /q) -1 \right) \right).
\end{equation}
If  $|\lr{rc}{q}| \leq r/2$, the error term in the above expression is easily
seen to be bounded by $O(1/q)$.  We now show that if $|\lr{rc}{q}| \leq r/2$,
the above integral is large,
so the probability of obtaining a state $\|c,x^k\mod{n}\>$ is large.
Note that this condition depends only on $c$ and is independent of~$k$.
Substituting $u=rb/q$ in the above integral,
we get
\begin{equation}
\frac{1}{r} \int_{0}^{\frac{r}{q}\left\lfloor\frac{q-k-1}{r}\right\rfloor}
\exp\left( 2\pi i {\textstyle\frac{\lr{rc}{q}}{r}}u \right) du.
\end{equation}
Since $k < r$, approximating the upper limit of integration by~1 results 
in only a $O(1/q)$ 
error in the above expression.  If we do this, we obtain the integral
\begin{equation}
\frac{1}{r} \int_{0}^{1} 
\exp\left(2\pi i {\textstyle\frac{\lr{rc}{q}}{r}}u \right) du.
\label{factorint3}
\end{equation}
Letting $\lr{rc}{q}/r$ vary between $-\frac{1}{2}$ and $\frac{1}{2}$,
the absolute magnitude of the integral (\ref{factorint3}) is easily seen
to be minimized when $\lr{rc}{q}/{r} = \pm \frac{1}{2}$, in which case 
the absolute value of expression (\ref{factorint3}) is $2/(\pi r)$.
The square of this quantity is a lower bound on
the probability that we see any particular 
state $\|c, x^k\mod{n}\>$ with $\lr{rc}{q} \leq r/2$;
this probability is thus asymptotically
bounded below by $4/(\pi^2 r^2)$, and so is at
least $1/3r^2$ for sufficiently large~$n$.  


\begin{figure}
\begin{center}
\setlength{\unitlength}{0.240900pt}
\ifx\plotpoint\undefined\newsavebox{\plotpoint}\fi
\sbox{\plotpoint}{\rule[-0.175pt]{0.350pt}{0.350pt}}%
\begin{picture}(1500,900)(0,0)
\small
\sbox{\plotpoint}{\rule[-0.175pt]{0.350pt}{0.350pt}}%
\put(242,158){\makebox(0,0)[r]{0.00}}
\put(244,158){\rule[-0.175pt]{4.818pt}{0.350pt}}
\put(242,263){\makebox(0,0)[r]{0.02}}
\put(244,263){\rule[-0.175pt]{4.818pt}{0.350pt}}
\put(242,368){\makebox(0,0)[r]{0.04}}
\put(244,368){\rule[-0.175pt]{4.818pt}{0.350pt}}
\put(242,473){\makebox(0,0)[r]{0.06}}
\put(244,473){\rule[-0.175pt]{4.818pt}{0.350pt}}
\put(242,577){\makebox(0,0)[r]{0.08}}
\put(244,577){\rule[-0.175pt]{4.818pt}{0.350pt}}
\put(242,682){\makebox(0,0)[r]{0.10}}
\put(244,682){\rule[-0.175pt]{4.818pt}{0.350pt}}
\put(242,787){\makebox(0,0)[r]{0.12}}
\put(244,787){\rule[-0.175pt]{4.818pt}{0.350pt}}
\put(329,113){\makebox(0,0){0}}
\put(329,138){\rule[-0.175pt]{0.350pt}{4.818pt}}
\put(459,113){\makebox(0,0){32}}
\put(459,138){\rule[-0.175pt]{0.350pt}{4.818pt}}
\put(590,113){\makebox(0,0){64}}
\put(590,138){\rule[-0.175pt]{0.350pt}{4.818pt}}
\put(720,113){\makebox(0,0){96}}
\put(720,138){\rule[-0.175pt]{0.350pt}{4.818pt}}
\put(850,113){\makebox(0,0){128}}
\put(850,138){\rule[-0.175pt]{0.350pt}{4.818pt}}
\put(980,113){\makebox(0,0){160}}
\put(980,138){\rule[-0.175pt]{0.350pt}{4.818pt}}
\put(1110,113){\makebox(0,0){192}}
\put(1110,138){\rule[-0.175pt]{0.350pt}{4.818pt}}
\put(1241,113){\makebox(0,0){224}}
\put(1241,138){\rule[-0.175pt]{0.350pt}{4.818pt}}
\put(1371,113){\makebox(0,0){256}}
\put(1371,138){\rule[-0.175pt]{0.350pt}{4.818pt}}
\put(264,158){\rule[-0.175pt]{282.335pt}{0.350pt}}
\put(1436,158){\rule[-0.175pt]{0.350pt}{151.526pt}}
\put(264,787){\rule[-0.175pt]{282.335pt}{0.350pt}}
\put(45,472){\makebox(0,0)[l]{\shortstack{P}}}
\put(850,68){\makebox(0,0){\it c}}
\put(264,158){\rule[-0.175pt]{0.350pt}{151.526pt}}
\put(329,158){\rule[-0.175pt]{0.350pt}{126.232pt}}
\put(333,158){\usebox{\plotpoint}}
\put(337,158){\usebox{\plotpoint}}
\put(341,158){\usebox{\plotpoint}}
\put(345,158){\usebox{\plotpoint}}
\put(349,158){\usebox{\plotpoint}}
\put(354,158){\usebox{\plotpoint}}
\put(358,158){\usebox{\plotpoint}}
\put(362,158){\usebox{\plotpoint}}
\put(366,158){\usebox{\plotpoint}}
\put(370,158){\usebox{\plotpoint}}
\put(374,158){\usebox{\plotpoint}}
\put(378,158){\usebox{\plotpoint}}
\put(382,158){\usebox{\plotpoint}}
\put(386,158){\usebox{\plotpoint}}
\put(390,158){\usebox{\plotpoint}}
\put(394,158){\usebox{\plotpoint}}
\put(398,158){\usebox{\plotpoint}}
\put(402,158){\usebox{\plotpoint}}
\put(406,158){\usebox{\plotpoint}}
\put(411,158){\rule[-0.175pt]{0.350pt}{0.482pt}}
\put(415,158){\rule[-0.175pt]{0.350pt}{0.482pt}}
\put(419,158){\rule[-0.175pt]{0.350pt}{0.964pt}}
\put(423,158){\rule[-0.175pt]{0.350pt}{1.686pt}}
\put(427,158){\rule[-0.175pt]{0.350pt}{4.577pt}}
\put(431,158){\rule[-0.175pt]{0.350pt}{32.281pt}}
\put(435,158){\rule[-0.175pt]{0.350pt}{72.270pt}}
\put(439,158){\rule[-0.175pt]{0.350pt}{6.022pt}}
\put(443,158){\rule[-0.175pt]{0.350pt}{1.927pt}}
\put(447,158){\rule[-0.175pt]{0.350pt}{0.964pt}}
\put(451,158){\rule[-0.175pt]{0.350pt}{0.723pt}}
\put(455,158){\rule[-0.175pt]{0.350pt}{0.482pt}}
\put(459,158){\usebox{\plotpoint}}
\put(463,158){\usebox{\plotpoint}}
\put(467,158){\usebox{\plotpoint}}
\put(472,158){\usebox{\plotpoint}}
\put(476,158){\usebox{\plotpoint}}
\put(480,158){\usebox{\plotpoint}}
\put(484,158){\usebox{\plotpoint}}
\put(488,158){\usebox{\plotpoint}}
\put(492,158){\usebox{\plotpoint}}
\put(496,158){\usebox{\plotpoint}}
\put(500,158){\usebox{\plotpoint}}
\put(504,158){\usebox{\plotpoint}}
\put(508,158){\usebox{\plotpoint}}
\put(512,158){\usebox{\plotpoint}}
\put(516,158){\usebox{\plotpoint}}
\put(520,158){\usebox{\plotpoint}}
\put(524,158){\rule[-0.175pt]{0.350pt}{0.482pt}}
\put(529,158){\rule[-0.175pt]{0.350pt}{0.964pt}}
\put(533,158){\rule[-0.175pt]{0.350pt}{3.132pt}}
\put(537,158){\rule[-0.175pt]{0.350pt}{110.573pt}}
\put(541,158){\rule[-0.175pt]{0.350pt}{6.986pt}}
\put(545,158){\rule[-0.175pt]{0.350pt}{1.445pt}}
\put(549,158){\rule[-0.175pt]{0.350pt}{0.482pt}}
\put(553,158){\usebox{\plotpoint}}
\put(557,158){\usebox{\plotpoint}}
\put(561,158){\usebox{\plotpoint}}
\put(565,158){\usebox{\plotpoint}}
\put(569,158){\usebox{\plotpoint}}
\put(573,158){\usebox{\plotpoint}}
\put(577,158){\usebox{\plotpoint}}
\put(581,158){\usebox{\plotpoint}}
\put(585,158){\usebox{\plotpoint}}
\put(590,158){\usebox{\plotpoint}}
\put(594,158){\usebox{\plotpoint}}
\put(598,158){\usebox{\plotpoint}}
\put(602,158){\usebox{\plotpoint}}
\put(606,158){\usebox{\plotpoint}}
\put(610,158){\usebox{\plotpoint}}
\put(614,158){\usebox{\plotpoint}}
\put(618,158){\usebox{\plotpoint}}
\put(622,158){\usebox{\plotpoint}}
\put(626,158){\usebox{\plotpoint}}
\put(630,158){\rule[-0.175pt]{0.350pt}{0.482pt}}
\put(634,158){\rule[-0.175pt]{0.350pt}{1.445pt}}
\put(638,158){\rule[-0.175pt]{0.350pt}{6.986pt}}
\put(642,158){\rule[-0.175pt]{0.350pt}{110.573pt}}
\put(647,158){\rule[-0.175pt]{0.350pt}{3.132pt}}
\put(651,158){\rule[-0.175pt]{0.350pt}{0.964pt}}
\put(655,158){\rule[-0.175pt]{0.350pt}{0.482pt}}
\put(659,158){\usebox{\plotpoint}}
\put(663,158){\usebox{\plotpoint}}
\put(667,158){\usebox{\plotpoint}}
\put(671,158){\usebox{\plotpoint}}
\put(675,158){\usebox{\plotpoint}}
\put(679,158){\usebox{\plotpoint}}
\put(683,158){\usebox{\plotpoint}}
\put(687,158){\usebox{\plotpoint}}
\put(691,158){\usebox{\plotpoint}}
\put(695,158){\usebox{\plotpoint}}
\put(699,158){\usebox{\plotpoint}}
\put(704,158){\usebox{\plotpoint}}
\put(708,158){\usebox{\plotpoint}}
\put(712,158){\usebox{\plotpoint}}
\put(716,158){\usebox{\plotpoint}}
\put(720,158){\usebox{\plotpoint}}
\put(724,158){\rule[-0.175pt]{0.350pt}{0.482pt}}
\put(728,158){\rule[-0.175pt]{0.350pt}{0.723pt}}
\put(732,158){\rule[-0.175pt]{0.350pt}{0.964pt}}
\put(736,158){\rule[-0.175pt]{0.350pt}{1.927pt}}
\put(740,158){\rule[-0.175pt]{0.350pt}{6.022pt}}
\put(744,158){\rule[-0.175pt]{0.350pt}{72.270pt}}
\put(748,158){\rule[-0.175pt]{0.350pt}{32.281pt}}
\put(752,158){\rule[-0.175pt]{0.350pt}{4.577pt}}
\put(756,158){\rule[-0.175pt]{0.350pt}{1.686pt}}
\put(760,158){\rule[-0.175pt]{0.350pt}{0.964pt}}
\put(765,158){\rule[-0.175pt]{0.350pt}{0.482pt}}
\put(769,158){\rule[-0.175pt]{0.350pt}{0.482pt}}
\put(773,158){\usebox{\plotpoint}}
\put(777,158){\usebox{\plotpoint}}
\put(781,158){\usebox{\plotpoint}}
\put(785,158){\usebox{\plotpoint}}
\put(789,158){\usebox{\plotpoint}}
\put(793,158){\usebox{\plotpoint}}
\put(797,158){\usebox{\plotpoint}}
\put(801,158){\usebox{\plotpoint}}
\put(805,158){\usebox{\plotpoint}}
\put(809,158){\usebox{\plotpoint}}
\put(813,158){\usebox{\plotpoint}}
\put(817,158){\usebox{\plotpoint}}
\put(822,158){\usebox{\plotpoint}}
\put(826,158){\usebox{\plotpoint}}
\put(830,158){\usebox{\plotpoint}}
\put(834,158){\usebox{\plotpoint}}
\put(838,158){\usebox{\plotpoint}}
\put(842,158){\usebox{\plotpoint}}
\put(846,158){\usebox{\plotpoint}}
\put(850,158){\rule[-0.175pt]{0.350pt}{126.232pt}}
\put(854,158){\usebox{\plotpoint}}
\put(858,158){\usebox{\plotpoint}}
\put(862,158){\usebox{\plotpoint}}
\put(866,158){\usebox{\plotpoint}}
\put(870,158){\usebox{\plotpoint}}
\put(874,158){\usebox{\plotpoint}}
\put(878,158){\usebox{\plotpoint}}
\put(883,158){\usebox{\plotpoint}}
\put(887,158){\usebox{\plotpoint}}
\put(891,158){\usebox{\plotpoint}}
\put(895,158){\usebox{\plotpoint}}
\put(899,158){\usebox{\plotpoint}}
\put(903,158){\usebox{\plotpoint}}
\put(907,158){\usebox{\plotpoint}}
\put(911,158){\usebox{\plotpoint}}
\put(915,158){\usebox{\plotpoint}}
\put(919,158){\usebox{\plotpoint}}
\put(923,158){\usebox{\plotpoint}}
\put(927,158){\usebox{\plotpoint}}
\put(931,158){\rule[-0.175pt]{0.350pt}{0.482pt}}
\put(935,158){\rule[-0.175pt]{0.350pt}{0.482pt}}
\put(940,158){\rule[-0.175pt]{0.350pt}{0.964pt}}
\put(944,158){\rule[-0.175pt]{0.350pt}{1.686pt}}
\put(948,158){\rule[-0.175pt]{0.350pt}{4.577pt}}
\put(952,158){\rule[-0.175pt]{0.350pt}{32.281pt}}
\put(956,158){\rule[-0.175pt]{0.350pt}{72.270pt}}
\put(960,158){\rule[-0.175pt]{0.350pt}{6.022pt}}
\put(964,158){\rule[-0.175pt]{0.350pt}{1.927pt}}
\put(968,158){\rule[-0.175pt]{0.350pt}{0.964pt}}
\put(972,158){\rule[-0.175pt]{0.350pt}{0.723pt}}
\put(976,158){\rule[-0.175pt]{0.350pt}{0.482pt}}
\put(980,158){\usebox{\plotpoint}}
\put(984,158){\usebox{\plotpoint}}
\put(988,158){\usebox{\plotpoint}}
\put(992,158){\usebox{\plotpoint}}
\put(997,158){\usebox{\plotpoint}}
\put(1001,158){\usebox{\plotpoint}}
\put(1005,158){\usebox{\plotpoint}}
\put(1009,158){\usebox{\plotpoint}}
\put(1013,158){\usebox{\plotpoint}}
\put(1017,158){\usebox{\plotpoint}}
\put(1021,158){\usebox{\plotpoint}}
\put(1025,158){\usebox{\plotpoint}}
\put(1029,158){\usebox{\plotpoint}}
\put(1033,158){\usebox{\plotpoint}}
\put(1037,158){\usebox{\plotpoint}}
\put(1041,158){\usebox{\plotpoint}}
\put(1045,158){\rule[-0.175pt]{0.350pt}{0.482pt}}
\put(1049,158){\rule[-0.175pt]{0.350pt}{0.964pt}}
\put(1053,158){\rule[-0.175pt]{0.350pt}{3.132pt}}
\put(1058,158){\rule[-0.175pt]{0.350pt}{110.573pt}}
\put(1062,158){\rule[-0.175pt]{0.350pt}{6.986pt}}
\put(1066,158){\rule[-0.175pt]{0.350pt}{1.445pt}}
\put(1070,158){\rule[-0.175pt]{0.350pt}{0.482pt}}
\put(1074,158){\usebox{\plotpoint}}
\put(1078,158){\usebox{\plotpoint}}
\put(1082,158){\usebox{\plotpoint}}
\put(1086,158){\usebox{\plotpoint}}
\put(1090,158){\usebox{\plotpoint}}
\put(1094,158){\usebox{\plotpoint}}
\put(1098,158){\usebox{\plotpoint}}
\put(1102,158){\usebox{\plotpoint}}
\put(1106,158){\usebox{\plotpoint}}
\put(1110,158){\usebox{\plotpoint}}
\put(1115,158){\usebox{\plotpoint}}
\put(1119,158){\usebox{\plotpoint}}
\put(1123,158){\usebox{\plotpoint}}
\put(1127,158){\usebox{\plotpoint}}
\put(1131,158){\usebox{\plotpoint}}
\put(1135,158){\usebox{\plotpoint}}
\put(1139,158){\usebox{\plotpoint}}
\put(1143,158){\usebox{\plotpoint}}
\put(1147,158){\usebox{\plotpoint}}
\put(1151,158){\rule[-0.175pt]{0.350pt}{0.482pt}}
\put(1155,158){\rule[-0.175pt]{0.350pt}{1.445pt}}
\put(1159,158){\rule[-0.175pt]{0.350pt}{6.986pt}}
\put(1163,158){\rule[-0.175pt]{0.350pt}{110.573pt}}
\put(1167,158){\rule[-0.175pt]{0.350pt}{3.132pt}}
\put(1171,158){\rule[-0.175pt]{0.350pt}{0.964pt}}
\put(1176,158){\rule[-0.175pt]{0.350pt}{0.482pt}}
\put(1180,158){\usebox{\plotpoint}}
\put(1184,158){\usebox{\plotpoint}}
\put(1188,158){\usebox{\plotpoint}}
\put(1192,158){\usebox{\plotpoint}}
\put(1196,158){\usebox{\plotpoint}}
\put(1200,158){\usebox{\plotpoint}}
\put(1204,158){\usebox{\plotpoint}}
\put(1208,158){\usebox{\plotpoint}}
\put(1212,158){\usebox{\plotpoint}}
\put(1216,158){\usebox{\plotpoint}}
\put(1220,158){\usebox{\plotpoint}}
\put(1224,158){\usebox{\plotpoint}}
\put(1228,158){\usebox{\plotpoint}}
\put(1233,158){\usebox{\plotpoint}}
\put(1237,158){\usebox{\plotpoint}}
\put(1241,158){\usebox{\plotpoint}}
\put(1245,158){\rule[-0.175pt]{0.350pt}{0.482pt}}
\put(1249,158){\rule[-0.175pt]{0.350pt}{0.723pt}}
\put(1253,158){\rule[-0.175pt]{0.350pt}{0.964pt}}
\put(1257,158){\rule[-0.175pt]{0.350pt}{1.927pt}}
\put(1261,158){\rule[-0.175pt]{0.350pt}{6.022pt}}
\put(1265,158){\rule[-0.175pt]{0.350pt}{72.270pt}}
\put(1269,158){\rule[-0.175pt]{0.350pt}{32.281pt}}
\put(1273,158){\rule[-0.175pt]{0.350pt}{4.577pt}}
\put(1277,158){\rule[-0.175pt]{0.350pt}{1.686pt}}
\put(1281,158){\rule[-0.175pt]{0.350pt}{0.964pt}}
\put(1285,158){\rule[-0.175pt]{0.350pt}{0.482pt}}
\put(1290,158){\rule[-0.175pt]{0.350pt}{0.482pt}}
\put(1294,158){\usebox{\plotpoint}}
\put(1298,158){\usebox{\plotpoint}}
\put(1302,158){\usebox{\plotpoint}}
\put(1306,158){\usebox{\plotpoint}}
\put(1310,158){\usebox{\plotpoint}}
\put(1314,158){\usebox{\plotpoint}}
\put(1318,158){\usebox{\plotpoint}}
\put(1322,158){\usebox{\plotpoint}}
\put(1326,158){\usebox{\plotpoint}}
\put(1330,158){\usebox{\plotpoint}}
\put(1334,158){\usebox{\plotpoint}}
\put(1338,158){\usebox{\plotpoint}}
\put(1342,158){\usebox{\plotpoint}}
\put(1346,158){\usebox{\plotpoint}}
\put(1351,158){\usebox{\plotpoint}}
\put(1355,158){\usebox{\plotpoint}}
\put(1359,158){\usebox{\plotpoint}}
\put(1363,158){\usebox{\plotpoint}}
\put(1367,158){\usebox{\plotpoint}}
\end{picture}
\end{center}
\caption{The probability $\rm P$ of observing values of $c$ between 
$0$ and $255$, given
$q=256$ and $r=10$.}
\label{probabilityplot}
\end{figure}
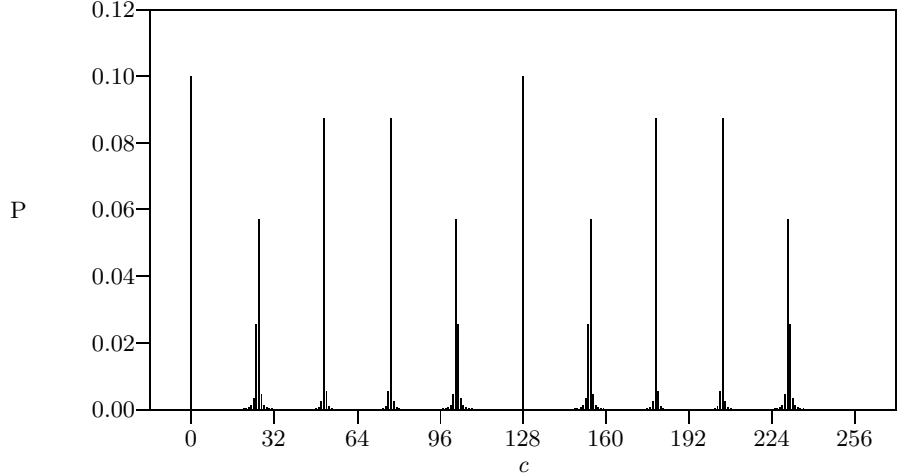

The probability of seeing a given
state $\|c,x^k\mod{n}\>$ will thus be at least $1/3r^2$ if 
\begin{equation} 
\frac{-r}{2} \leq \lr{rc}{q} \leq \frac{r}{2}, 
\end{equation}
i.e., if there is a $d$ such that
\begin{equation} \frac{-r}{2} \leq rc-dq \leq \frac{r}{2}. \end{equation}
Dividing by $rq$ and rearranging the terms gives
\begin{equation}
\left| \frac{c}{q} - \frac{d}{r} \right| \leq \frac{1}{2q}. 
\end{equation}
We know $c$ and~$q$.  
Because $q > n^2$, there is at most one fraction $d/r$ with $r<n$
that satisfies the above inequality.  
Thus, we can obtain the fraction $d/r$ in lowest terms by rounding 
$c/q$ to the nearest fraction having a denominator smaller than~$n$.
This fraction can be found in polynomial time by using a continued fraction 
expansion of $c/q$, which finds all the best approximations of 
$c/q$ by fractions [Hardy and Wright 1979, Chapter X, Knuth 1981].

The exact probabilities as 
given by equation (\ref{cprobs}) for an example case with $r=10$ and
$q=256$ are plotted in 
Figure~\ref{probabilityplot}.  The value $r=10$ could occur when factoring
33 if $x$ were chosen to be~5, for example. Here $q$ is taken smaller than 
$33^2$ so as to make the values of $c$ in the plot distinguishable;
this does not change the functional structure of ${\rm P}(c)$.  
Note that with high probability the
observed value of $c$ is near an integral multiple of $q/r = 256/10$.

If we have the fraction $d/r$ in lowest terms, and if $d$ happens to be
relatively prime to~$r$, this will give us~$r$.
We will now count the number of states $\|c,x^k\mod{n}\>$ which
enable us to compute $r$ in this way.  There are $\phi(r)$ possible values 
of $d$ relatively prime to~$r$, where $\phi$ is Euler's totient function
[Knuth 1981, Hardy and Wright 1979, \S5.5].
Each of these fractions $d/r$ is close to one fraction $c/q$ with
$|c/q-d/r| \leq 1/2q$.  There are also $r$ possible values for~$x^k$,
since $r$ is the order of~$x$.  Thus, there are $r \phi(r)$ states
$\|c,x^k\mod{n}\>$ which would enable us to obtain~$r$.  Since each
of these states occurs with probability at least $1/3r^2$, we obtain
$r$ with probability at least $\phi(r)/3r$.  Using the theorem that
$\phi(r)/r > \delta/\log \log r$ for some constant $\delta$ 
[Hardy and Wright 1979, Theorem 328], this shows that we find $r$ at least a 
$\delta/\log\log r$ fraction of the time, so by repeating this experiment
only $O(\log\log r)$ times, we are assured of a high probability of 
success.

In practice, assuming that quantum computation is more expensive than
classical computation, it would be worthwhile to alter the above algorithm
so as to perform less quantum computation and
more postprocessing.  First, if the observed state is $\|c\>$,
it would be wise to also try numbers close to $c$ such as $c\pm 1$, $c\pm 2$,
$\ldots$, since these also have a reasonable chance of being close to a
fraction $qd/r$.  Second, if $c/q \approx d/r$, and $d$ and $r$ have a 
common factor, it is likely to be small.  Thus,
if the observed value of $c/q$ is rounded off to
$d'/r'$ in lowest terms, for a
candidate $r$ one should consider not only~$r'$ but also its small multiples
$2r'$, $3r'$, \ldots, to see if these are the actual order of~$x$.  
Although the first technique will only reduce the expected number of trials
required to find $r$ by a constant factor, 
the second technique will reduce the expected number of trials 
for the hardest $n$ from $O(\log \log n)$ to $O(1)$ 
if the first $(\log n)^{1+\epsilon}$ multiples of~$r'$ are considered
\cite{Odly}.  A third technique is, if 
two candidate $r$'s have been found, say
$r_1$ and~$r_2$, to test the least common multiple of $r_1$ and $r_2$
as a candidate~$r$.  This third technique is also able to reduce the
expected number of trials to a constant \cite{Knil}, and will also work in 
some cases where the first two techniques fail.  

Note that in this algorithm for determining the order of an
element, we did not use many of the 
properties of multiplication $\bmod{n}$.  In fact, if we have a permutation
$f$ mapping the set $\{0,1,2,\ldots,n-1\}$ into itself such
that its $k$th iterate, $f^{(k)}(a)$, is computable in time polynomial 
in $\log n$ and $\log k$,
the same algorithm will be able to find the order of an element $a$ 
under~$f$, i.e., the minimum $r$ such that $f^{(r)}(a)=a$.

\section{Discrete logarithms}
For every prime~$p$, the multiplicative group $\bmod{p}$ is cyclic,
that is, there are generators $g$ such that $1$, $g$, $g^2$, \ldots, 
$g^{p-2}$
comprise all the non-zero residues $\bmod{p}$ [Hardy and Wright 1979,
Theorem 111, Knuth 1981].  Suppose we are given a prime $p$ and such a 
generator~$g$.  The {\em discrete logarithm} of a number $x$ with
respect to $p$ and $g$ is the integer $r$ with $0 \leq r < p-1$ such that
$g^r \equiv x \mod{p}$.  The fastest algorithm known for finding discrete
logarithms modulo arbitrary primes $p$ is Gordon's [1993] adaptation of 
the number field sieve, which runs in time 
$\exp(O(\log p)^{1/3} (\log \log p)^{2/3}))$.  We show how to find discrete
logarithms on a quantum computer with two 
modular exponentiations and two quantum Fourier transforms.  

This algorithm will use three quantum registers.  We first
find $q$ a power of 2 such that $q$ is close to~$p$, i.e., with
$p < q < 2p$.
Next, we put the first two registers in 
our quantum computer in the uniform superposition of
all $\|a\>$ and $\|b\>$ 
$\bmod{p-1}$, and compute $g^ax^{-b}\mod{p}$ in the third register.  This 
leaves our machine in the state
\begin{equation}
\frac{1}{p-1}\sum_{a=0}^{p-2}\sum_{b=0}^{p-2} \|a,b,g^a x^{-b}\mod{p}\>.
\end{equation}
As before, we use the Fourier transform $A_q$ to
send $\|a\>\rightarrow \|c\>$ and $\|b\>\rightarrow \|d\>$
with probability amplitude $\frac{1}{q}\exp(2\pi i (ac+bd)/q)$.
This is, we take the state $\|a,b\>$ to the state
\begin{equation}
\frac{1}{q} \sum_{c=0}^{q-1} \sum_{d=0}^{q-1} 
\textstyle \exp\big(\frac{2 \pi i}{q} (ac+bd)\big)
\|c,d\>.
\end{equation}
This leaves our quantum computer in the state
\begin{equation}
{\frac{1}{(p-1)q}}
\sum_{a,b=0\ }^{p-2} \sum_{\ c,d=0}^{q-1}
\textstyle \exp\big(\frac{2 \pi i}{q} (ac+bd)\big)
\|c,d,g^a x^{-b}\!\mod{p}\>.
\end{equation}
Finally, we observe the state of the quantum computer.

The probability of observing a state $\|c,d,y\>$ with 
$y \equiv g^k \mod{p}$ is
\begin{equation}
\left| \; \frac{1}{(p-1)q}\sum_{{a,b}\atop{a-rb \equiv k}}
\exp\left({\textstyle \frac{2\pi i}{q} } (ac+bd)\right) \;
\right|^2
\label{probobserve}
\end{equation}
where the sum is over all
$(a,b)$ such that $a-rb \equiv k \mod{p-1}$.
Note that we now have two moduli to deal with, $p-1$ and~$q$.  While this
makes keeping track of things more confusing, it does not pose serious
problems.
We now use the relation
\begin{equation}
a = br + k - (p-1) \left\lfloor{\textstyle\frac{br+k}{p-1}}\right\rfloor
\label{arelation}
\end{equation}
and substitute (\ref{arelation}) in the expression (\ref{probobserve})
to obtain the amplitude
on $\|c,d,g^k\mod{p}\>$, which is
\begin{equation}
{\frac{1}{(p-1)q}}
\sum_{b=0}^{p-2}
\exp\Big(\textstyle\frac{2\pi i}{q} \big(brc+kc+bd-c(p-1)\left\lfloor
{\textstyle\frac{br+k}{p-1}}\right\rfloor\big)\Big).
\label{amplitudeexpress}
\end{equation}
The absolute value of the square of this amplitude is the probability
of observing the state $\|c,d,g^k\mod{p}\>$.
We will now analyze the expression 
(\ref{amplitudeexpress}).  First, a factor of 
$\exp(2\pi i k c/q)$ can be taken out of all the terms and ignored, 
because it does not change the probability.  Next, we split the
exponent into two parts and factor out $b$ to obtain
\begin{equation}
\frac{1}{(p-1)q}\sum_{b=0}^{p-2}
\textstyle
\exp\left(\frac{2\pi i}{q} 
bT\right)
\exp\left(\frac{2\pi i}{q} 
V\right),
\label{bigeq}
\end{equation}
where
\begin{equation}
T = \textstyle  {rc+d-\frac{r}{p-1}\lr{c(p-1)}{q}},\!\phantom{pen}
\end{equation}
and
\begin{equation}
V = \textstyle{\left(\frac{br}{p-1}-
\left\lfloor\frac{br+k}{p-1}\right\rfloor\right)\lr{c(p-1)}{q}}.
\end{equation}
Here by $\lr{z}{q}$ we mean the residue of $z \mod q$ with
$-q/2 < \lr{z}{q} \leq q/2$, as in equation (\ref{cprobs}).

We next classify possible outputs (observed states) of the quantum
computer into ``good'' and ``bad.''
We will show that if we get enough ``good'' outputs, then we will likely
be able to
deduce~$r$, and that furthermore, the chance of getting a ``good'' 
output is constant.  The idea is that if 
\begin{equation}
\big|\lr{T}{q}\big| =
\big| rc+d-{\textstyle \frac{r}{p-1}}\lr{c(p-1)}{q} -jq \big| 
\leq \frac{1}{2},
\label{firstcond}
\end{equation}
where $j$ is the closest integer to $ T/q $,
then as $b$ varies between 0 and $p-2$, the phase of 
the first exponential term in equation (\ref{bigeq}) only varies 
over at most half of the unit circle.  Further, if 
\begin{equation}
\left| \lr{c(p-1)}{q} \right| \leq q/12,
\label{secondcond}
\end{equation}
then $|V|$ is always at most $q/12$, so the phase of
the second exponential term in equation (\ref{bigeq}) never is farther than
$\exp(\pi i/ 6)$ from~1.  If conditions (\ref{firstcond}) and 
(\ref{secondcond}) both hold, we will say that an output is
``good.''  We will
show that if both conditions hold, then the contribution to the
probability from the corresponding term is significant.  Furthermore, 
both conditions will hold with constant probability, and
a reasonable sample of $c$'s for which condition (\ref{firstcond}) holds
will allow us to deduce~$r$.  

We now give a lower bound
on the probability of each good output, i.e., an output that satisfies 
conditions (\ref{firstcond}) and (\ref{secondcond}).  
We know that as $b$ ranges from 0 to $p-2$, the phase of
$\exp(2\pi i bT/q)$ ranges from
$0$ to $2\pi i W$ where
\begin{equation}
W = \frac{p-2}{q} 
{\left(rc+d-\frac{r}{p-1}\lr{c(p-1)}{q} -jq \right)}
\end{equation}
and $j$ is as in equation (\ref{firstcond}).
Thus, the component of the
amplitude of the first exponential in the summand of (\ref{bigeq}) 
in the direction 
\begin{equation}
\exp\left( \pi i W
\right)
\label{direction}
\end{equation}
is at least $\cos(2 \pi \left| W/2 - Wb/(p-2)\right|)$.  By condition 
(\ref{secondcond}), the phase can vary by at most ${\pi i/6}$ due to 
the second exponential $\exp(2\pi i V/q)$.  
Applying this variation in the manner that minimizes the component 
in the direction (\ref{direction}), we get that the component in this 
direction is at least 
\begin{equation}
\cos(2 \pi \left| W/2 - Wb/(p-2)\right| + {\textstyle\frac{\pi}{6}}).  
\end{equation}
Thus we get that the absolute value of the amplitude (\ref{bigeq}) is at 
least
\begin{equation}
\frac{1}{(p-1)q} \sum_{b=0}^{p-2} \cos\left(2 \pi \left| W/2-Wb/(p-2)\right| 
+ {\textstyle\frac{\pi}{6}}\right).
\end{equation}
Replacing this sum with an integral, we get that 
the absolute value of this amplitude is at least
\begin{equation}
\frac{2}{q} \int_{0}^{1/2} \cos (\textstyle \frac{\pi}{6} + 2\pi |W| u) du
\ + \ O\left(\frac{W}{pq}\right).
\label{logintegral1}
\end{equation}
 From condition (\ref{firstcond}), $|W| \leq \frac{1}{2}$, so the error term
is $O(\frac{1}{pq})$.
As $W$ varies between $-\frac{1}{2}$ and $\frac{1}{2}$, 
the integral (\ref{logintegral1})
is minimized when $|W|=\frac{1}{2}$.
Thus, the probability of 
arriving at a state $\|c,d,y\>$ that satisfies both conditions
(\ref{firstcond}) and (\ref{secondcond}) is at least
\begin{equation}
\left(
\frac{1}{q} \frac{2}{\pi} \int_{\pi/6}^{2\pi/3} \cos u \; du 
\right)^2,
\end{equation}
or at least $.054 / q^2 > 1/(20q^2)$.  

We will now count the number of pairs $(c,d)$ satisfying conditions
(\ref{firstcond}) and (\ref{secondcond}).
The number of pairs $(c,d)$ such
that (\ref{firstcond}) holds is exactly the number of possible 
$c$'s, since 
for every $c$ there is exactly one $d$ such that (\ref{firstcond})
holds.  Unless $\gcd(p-1,q)$ is large, the number of $c$'s
for which (\ref{secondcond}) holds is approximately $q/6$, and even if
it is large, this number is at least $q/12$.  
Thus, 
there are at least $q/12$ pairs $(c,d)$ satisfying 
both conditions.  Multiplying by $p-1$,
which is the number of possible $y$'s, gives approximately
$pq/12$ good states $\|c,d,y\>$.  Combining this calculation with the lower
bound $1/(20q^2)$ on the probability of observing
each good state gives us that the 
probability of observing some good state is at least $p/(240q)$, or at 
least $1/480$ (since $q < 2p$).  Note that each good
$c$ has a probability of at least $(p-1)/(20q^2) \geq 1/(40q)$ of being 
observed, since there $p-1$ values of $y$ and one value of $d$ with
which $c$ can make a good state $\|c,d,y\>$.

We now want to recover $r$ from a pair $c,d$ such that 
\begin{equation}
-\frac{1}{2q} \leq 
\frac{d}{q} + r\left(\frac{c(p-1)-\lr{c(p-1)}{q}}{(p-1)q}\right) \leq
\frac{1}{2q} \ \ \ \  \mod{1},
\end{equation}
where this equation was obtained from condition (\ref{firstcond}) by 
dividing by~$q$.  The first thing to notice is that the multiplier on~$r$ 
is a fraction with denominator $p-1$, since $q$ evenly divides 
$c(p-1) - \lr{c(p-1)}{q}$.  Thus, we
need only round $d/q$ off to the nearest multiple of $1/(p-1)$ and divide
$\bmod{p-1}$ by the integer
\begin{equation}
c' = \frac{c(p-1)-\lr{c(p-1)}{q}}{q}
\label{definecprime}
\end{equation}
to find a candidate~$r$.  To show that the quantum calculation need
only be repeated a polynomial number of times to find the correct~$r$
requires only a few more details.  The problem is that we cannot
divide by a number $c'$ which is not relatively prime to $p-1$.

For the 
discrete log algorithm, we do not know that all possible values
of $c'$ are generated with reasonable likelihood; we only know this about
one-twelfth of them.  This additional difficulty makes the next step harder
than the corresponding step in the algorithm for factoring.  If we knew the
remainder of $r$ modulo all prime powers dividing $p-1$, we could use the
Chinese remainder theorem to recover $r$ in polynomial time.  We will only
be able to prove that we can find this remainder
for primes larger than~18, but with a little extra work
we will still be able to recover~$r$.

Recall that each good $(c,d)$ pair is generated with probability
at least $1/(20q^2)$, and that at least a twelfth of the possible $c$'s 
are in a good $(c,d)$ pair.  From equation (\ref{definecprime}), it follows
that these $c$'s are mapped from $c/q$ to
$c'/(p-1)$ by rounding to the nearest integral multiple of \linebreak[3]
$1/\linebreak[1](p-1)$.
Further, the good $c$'s are exactly those in which ${c}/{q}$ is
close to ${c'}/(p-1)$.  Thus, each good $c$ corresponds with exactly
one~$c'$.  We would like to show that for any prime power 
$p_i^{\alpha_i}$ dividing $p-1$, a random good $c'$ is unlikely to 
contain~$p_i$.  If we are willing to accept a large constant for 
our algorithm, we can just ignore the prime powers under~18; if we know 
$r$ modulo all prime powers
over~18, we can try all possible residues for primes under 18 with only a 
(large) constant factor increase in running time.   
Because at least one twelfth of the $c$'s were in a good
$(c,d)$ pair, at least one twelfth of the $c'$'s are good.  Thus,
for a prime power $p_i^{\alpha_i}$, a random good $c'$ is
divisible by $p_i^{\alpha_i}$ with probability at most $12/p_i^{\alpha_i}$.  
If we have $t$ good $c'$'s, the probability 
of having a prime power over 18 that divides all of them is therefore at most
\begin{equation}
\sum_{18 \,<\, p_i^{\alpha_i} \big| (p-1)} 
\left(\frac{12}{p_i^{\alpha_i}}\right)^t,
\end{equation}
where $a|b$ means that $a$ evenly divides~$b$, so
the sum is over all prime powers greater than 18 that divide $p-1$.
This sum (over all integers $> 18$) converges for $t=2$, and goes down
by at least a factor of $2/3$ for each further increase of $t$ by~1; thus for
some constant $t$ it is less than $1/2$.  

Recall that each good $c'$ 
is obtained with probability at least $1/(40q)$ from any experiment.  Since 
there are $q/12$ good $c'$'s, after $480t$ experiments, we are likely 
to obtain a sample of $t$ good $c'$'s chosen equally likely from all
good $c'$'s.  Thus, we will be able to find a set of $c'$'s such
that all prime powers $p_i^{\alpha_i} > 20$ dividing $p-1$ are relatively 
prime to at least one of these $c'$'s.  
To obtain a polynomial time algorithm,
all one need do is try all possible sets of $c'$'s of size~$t$; in practice,
one would use an algorithm to find sets of $c'$'s with large common factors.
This set gives the residue of $r$ for all primes larger than~18.  
For each prime $p_i$ less than~18, we have at most 18 possibilities
for the residue modulo $p_i^{\alpha_i}$, where $\alpha_i$ is the exponent
on prime $p_i$ in the prime factorization of $p-1$.  We can thus try all
possibilities for residues modulo powers of primes less than~18:  for each
possibility we can calculate the corresponding~$r$ using the Chinese 
remainder theorem and then check to see whether it is the desired discrete
logarithm.  

If one were to actually program this algorithm 
there are many ways in which the efficiency
could be increased over the efficiency shown in this paper.  
For example, the estimate for the number of good $c'$'s is likely too low, 
especially since weaker conditions than (\ref{firstcond}) and 
(\ref{secondcond}) should suffice.
This means that the number of times the experiment need be run could be
reduced.  It also
seems improbable that the distribution of bad values of $c'$ would have any 
relationship to primes under~18; if this is true,
we need not treat small prime powers separately.

This algorithm does not use very many properties of ${\rm Z}_p$, so we can
use the same algorithm to find discrete logarithms over other fields
such as $Z_{p^\alpha}$, as long as the field has a cyclic multiplicative 
group.   All we need is that we know the order of the 
generator, and that we can multiply and take inverses of elements in 
polynomial time.   The order of the generator could in fact be computed
using the quantum order-finding algorithm given in \S5 of this paper.
Boneh and Lipton [1995] have generalized the algorithm so as to be able to
find discrete logarithms when the group is abelian but not cyclic.

\section{Comments and open problems}
It is currently believed that the most difficult aspect
of building an actual quantum computer will be dealing with the problems of 
imprecision and decoherence.  It was shown by Bennett et al.
[1994] that the quantum gates need only have precision
$O(1/t)$ in order to have a reasonable probability of completing $t$ steps 
of quantum computation; that is, there is a $c$ such that
if the amplitudes in the unitary matrices
representing the quantum gates are all perturbed by at most $c/t$, the 
quantum computer will still have a reasonable chance of producing the 
desired output.  Similarly, the decoherence needs to be only polynomially 
small in $t$ in order to have a reasonable probability of completing 
$t$ steps of computation successfully.  This holds not only for the simple 
model of decoherence where each bit has a fixed probability of
decohering at each time step, but also for more complicated models of
decoherence which are derived from fundamental quantum mechanical 
considerations \cite{Unru, PaSuEk, ChLaShZu}.  However, building
quantum computers with high enough precision and low enough decoherence
to accurately perform long computations may present formidable 
difficulties to experimental physicists.  In classical computers, error 
probabilities can be reduced not only though hardware but also through 
software, by the use of redundancy and error-correcting codes.   The most 
obvious method of using redundancy in quantum computers is ruled out by 
the theorem that quantum bits cannot be cloned [Peres 1993, \S\mbox{9-4}], 
but this argument does
not rule out more complicated ways of reducing inaccuracy or decoherence
using software.  In fact, some progress in the direction of reducing
inaccuracy \cite{BeDeJo} and decoherence \cite{Shor2}
has already been made.
The result of Bennett et al.\ [1995] that quantum bits can
be faithfully transmitted over a noisy quantum channel gives further
hope that quantum computations can similarly be faithfully carried out 
using noisy quantum bits and noisy quantum gates.

Discrete logarithms and factoring are not in themselves widely useful
problems.  They have only become useful because they have been
found to be crucial for public-key cryptography, and this application
is in turn possible only because they have been presumed to be difficult.
This is also true of the generalizations of Boneh and Lipton [1995] of
these algorithms.
If the only uses of quantum computation remain discrete logarithms and 
factoring, it will likely become a
special-purpose technique whose only {\em raison d'\^etre}
is to thwart public key cryptosystems.  However, there may be other 
hard problems which could be solved asymptotically faster with quantum 
computers.  In particular, of interesting problems not known
to be NP-complete, the problem of finding a short vector in a lattice 
\cite{Adle,AdMc} seems as if it might potentially be 
amenable to solution by a quantum computer.  

In the history of computer science, however, most important problems
have turned out to be either polynomial-time or NP-complete.  Thus
quantum computers will likely not become widely useful unless they can 
solve NP-complete problems.  Solving NP-complete problems efficiently
is a Holy Grail of theoretical computer science which very few people expect
to be possible on a classical computer.  Finding polynomial-time algorithms
for solving these problems on a quantum computer would be a momentous
discovery.  There are some weak indications that
quantum computers are not powerful enough to solve NP-complete problems
\cite{BeBeBrVa}, 
but I do not believe that this potentiality should be ruled out as yet.

\section*{Acknowledgements}
I would like to thank Jeff Lagarias for finding and fixing a critical error
in the first version of the discrete log algorithm.  I would also like 
to thank him, David Applegate,
Charles Bennett, Gilles Brassard, Andrew Odlyzko, Dan Simon, 
Bob Solovay, Umesh Vazirani, and correspondents too numerous 
to list, for productive discussions, for corrections to and improvements of 
early drafts of this paper, and for pointers to the literature.

{

}

\end{document}